\documentclass[preprint,12pt]{elsarticle}

\usepackage{amssymb}

\usepackage{amsmath}
\usepackage{amsfonts}
\usepackage{latexsym}
\usepackage{graphicx}

\newcommand{\ra}{\rightarrow}
\newcommand{\bs}[1]{\boldsymbol{#1}}
\newcommand{\tn}[1]{\textnormal{#1}}
\newcommand{\mr}[1]{\mathrm{#1}}
\newcommand{\ap}[1]{\\approx}
\newcommand{\mb}[1]{\mbox{#1}}

\newcommand{\N}{\mathbb{N}}
\newcommand{\R}{\mathbb{R}}

\newcommand{\F}{\mathcal{F}}

\newtheorem{definition}{Definition}
\newtheorem{lemma}{Lemma}
\newtheorem{theorem}{Theorem}

\journal{Fuzzy Sets and Systems}
\begin{document}
\begin{frontmatter}


\title{Radial Fuzzy Systems}
\author{David Coufal}

\address{Institute of Computer Science, The Czech Academy of Sciences,\\
Pod Vod\'{a}renskou v\v{e}\v{z}\'{\i}~2, 182 07 Prague 8, Czech Republic}

\ead{david.coufal@cs.cas.cz}

\begin{abstract}
The class of radial fuzzy systems is introduced. The fuzzy systems in this class use radial functions
to implement membership functions of fuzzy sets and exhibit a shape preservation property
in antecedents of their rules. The property is called the radial property. It enables the radial
fuzzy systems to have their computational model mathematically tractable under both
conjunctive and implicative representations of their rule bases. Coherence of radial implicative
fuzzy systems is discussed and a sufficient condition for coherence is stated.
\end{abstract}

\begin{keyword}
fuzzy systems, radial functions, coherence



\end{keyword}

\end{frontmatter}


\section{Introduction}
\label{Sec01}
In applications of fuzzy computing, the notion of a fuzzy system plays a~key
role \citep{Handbook-GREEN, Wang97, Yager94, Klir95}. On a general level, 
the fuzzy system represents a function from an input set into an output set.
What is special about the fuzzy system is how this function is implemented.

The implementation draws on the four building blocks of the fuzzy system, 
namely: \textit{a fuzzifier}, \textit{a rule base}, \textit{an inference engine}
and \textit{a~defuzzifier}. The fuzzifier and defuzzifier are peripheral
blocks. The heart of the fuzzy system comprises the rule base, which stores
knowledge incorporated in the fuzzy system, and the inference engine
that constitutes a processing unit of the fuzzy system.

The knowledge stored in the rule base is canonically represented by a~set of 
IF-THEN \textit{rules} \citep{Dubois96, Handbook-VIOLET}. Each rule
consists of \textit{the antecedent} (the IF part) and \textit{the consequent}
(the THEN part). The antecedent and consequent are formed by fuzzy
sets that are specified on the input or the output set, respectively. The fuzzy sets 
enable the fuzzy system to incorporate information that is presented linguistically
in the way how people describe input-output relations. This ability is
the specialty that distinguishes the fuzzy systems from other modeling devices
such as neural networks and has brought success to them in several areas.
Among them, the area of process control is the most important one 
\citep{Handbook-GREEN, Wang97, Yager94, Driankov93}.

A concrete computational form of the function implemented by the fuzzy system
depends on a mathematical representation of the aforementioned building blocks.
A certain freedom is present here that leads to the different types of the fuzzy systems.
Diversity in shapes of the membership functions carries out a major portion
of this freedom. The other source of variability is the way how the rule base of
the fuzzy system is represented. The \textit{conjunctive} and the \textit{implicative}
representations are the most common~\citep{Dubois96, Handbook-VIOLET}.

Under the conjunctive representation, individual rules constitute characteristic
points in the input-output set. The whole rule base is then understood as the
list of these points and the fuzzy system interpolates between them during
its computation.



The implicative representation of the rule base is less frequent in applications, but
more challenging. In this case, each single rule is understood as an individual
condition/restriction imposed on a modeled input-output relation. Different rules
correspond to the different restrictions and are required to hold simultaneously.
The view of the stored knowledge is logically driven and the related fuzzy relation
can be seen as a kind of theory in the logical sense. The issue of \textit{coherence} 
(logical consistency) of the rule base is critical in this case \cite{Dubois97, Driankov93,
Stepnicka2013}.

Based on these facts, we are interested in the question of assessing the coherence 
of the implicative fuzzy systems. The assessment should be as comfortable as possible,
which means that we only want to check the parameters of the system.
We are further interested in a mutual relationship between the
implicative and conjunctive representations.

As it is common in science, a certain unifying view on phenomenon of interest is 
sought. However, a success is typically reached at the cost of a~certain simplification
and setting up assumptions. The task is to keep the extent of simplification on
that level that allows a researcher to theoretically tackle the problem, but still
to obtain non-trivial results and gain a broader view of the original problem.

The same approach is adopted it this paper. We restrict ourselves to the class of 
the so-called \textit{radial fuzzy systems}. In fact, we are going to introduce this
class here. The radial fuzzy systems use radial functions to implement fuzzy sets
in their rules and exhibit a shape preservation property in the antecedents of their
rules. This property enables the radial fuzzy systems to have a mathematically
tractable computational model. It is shown that the question of coherence can
be quite comfortably answered for this class. On the other hand, the restriction
to the radial fuzzy systems is not limiting because this class still contains a rich
variety of practically applicable systems including the most commonly used ones.

The paper is organized as follows. The next section discusses motivation for
introducing the radial fuzzy systems from the view of a practical
application. Sections 3 and 4 review the very basics of fuzzy set and systems 
theory in order to the paper be self-contained and the notions employed in
further sections have been clearly defined. Section 5 introduces the class 
of the radial fuzzy systems. Section~6 delivers computational models
of these systems. Section 7 deals with coherence of the radial implicative
fuzzy systems. Conclusions are given in Section 8 along with the directions
of future reasearch.

\section{Motivation}
\label{motiv}
When assessing computational capabilities of the fuzzy system, internal consistency
of its rule base is a highly desirable property. Consistency refers to the presence
of non-contradictory rules in the rule base. It is the \mbox{desirable} feature because
conflicting rules might have a fatal impact on the final computation of the system
as the following canonical example shows.

\begin{figure}[!htb]
\centering
\includegraphics[width=5.3in]{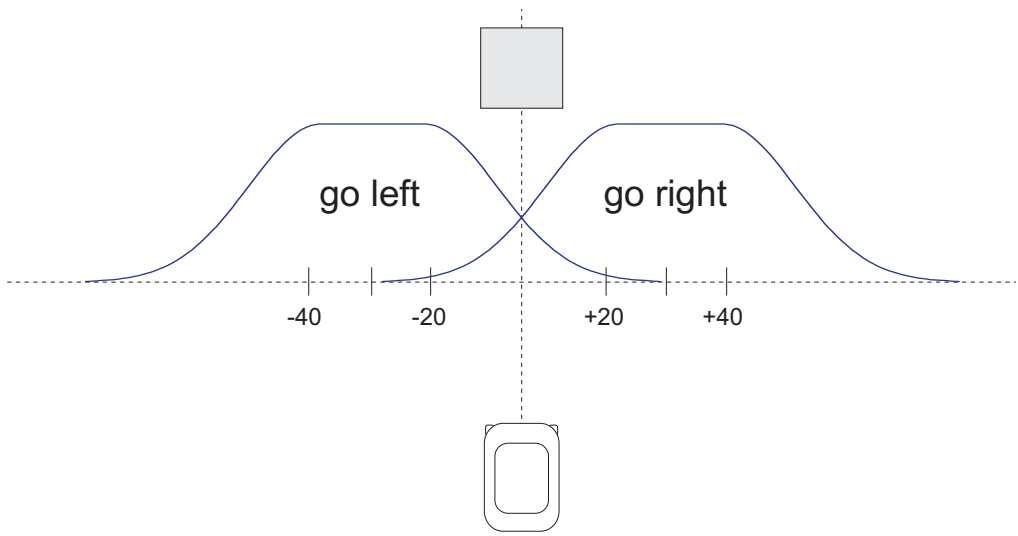}
\caption{Car navigation example.}
\label{fig_car}
\end{figure}

Suppose that the fuzzy system is used for car navigation and its rule base
consist only of two rules.  The input set is the set of possible locations
for a frontal obstacle that are measured in angles of deviation from the vertical axis.
The output set is the set of actions expressed as angles of future navigation. 
Both sets are equipped with the following fuzzy sets:  \textit{directly ahead} 
referring to angles about $0^{\circ}$, \textit{go left} to angles about
$-30^{\circ}$ and \textit{go right} to~angles about $+30^{\circ}$,
see Fig.~{\ref{fig_car}}. The rules are the following:
\begin{itemize}
\item
1.	IF \textit{an obstacle is directly ahead}  THEN \textit{go left} 
\item
2.	IF \textit{an obstacle is directly ahead}  THEN \textit{go right}
\end{itemize}

Suppose that the obstacle is located just at zero angle position, then both rules 
definitely fire.  Under the conjunctive representation of the rule base, the resulting
output fuzzy set before defuzzification has a camel-like shape due to its
bimodal character. Indeed, take two fuzzy sets from Fig.~{\ref{fig_car}}
and combine them by the maximum operator (or by another fuzzy disjunction). 
Applying the common MOM or COG defuzzification methods the fuzzy system
navigates the car just  to crash into the obstacle. This is the consequence
of inconsistency of the rules; irrespective of the fact that any of two rules, 
if applied alone, provides a good navigation policy.

In the case of the implicative representation, two fuzzy sets are combined
by a fuzzy conjunction, see Section \ref{ConImplSec} for details, and one is
interested in the core of the resulting output fuzzy set. In the above case, the core
is empty which indicates inconsistency. Indeed,  the left output fuzzy set
admits as fully compatible navigation the angles in range $[-40^{\circ},-20^{\circ}]$
and the other in range $[+20^{\circ},+40^{\circ}]$. These intervals are then combined
by the intersection which correspond to conjunctive combination in the computational
model of the implicative fuzzy systems. Hence the implicative systems have
capabilities to detect inconsistencies in their rule bases.

However, two issues arise here. One is that in today world of data abundance 
the rule bases are mostly created automatically by machine learning algorithms
from empirical data and detection of inconsistencies is forced to be performed 
on-line on the basis of characteristics of the created fuzzy systems.
Note that the data-drive rule bases consists of dozens or even hundreds of
rules and an expert assessment of consistency is impossible here. This
makes a serious problem especially when designing adaptive fuzzy controllers
\cite{Anstrom2008, Xajzl2013}.

The second issue deals with generalization of inconsistency detection
when going over the set of all possible inputs into the fuzzy system.
In~the above example, we have detected inconsistency for one fixed input 
($0^{\circ}$ angle location of the obstacle), but what about if the obstacle
was located at a different position, say $-30^{\circ}$. Would the two rules 
be still inconsistent? And what about all other possible locations of the obstacle? 
Generally speaking, we ask if for a given rule base and for any possible
input it can't happen that the rule base is inconsistent.  This quantification of
consistency over all possible inputs leads to the notion of coherence of 
the fuzzy implicative system \cite{Dubois97}.

In the paper, we show that both issues can be effectively handled for 
the class of the radial fuzzy systems, namely for the radial implicative
fuzzy systems. This makes a difference from ordinary fuzzy systems
where checking coherence leads to a somehow complicated and customized
procedures driven by the shape of employed fuzzy sets \cite{Dubois97, Leung93}.
In the case of the radial systems, a~unified view of the coherence checking
procedure is developed.

The radial conjunctive fuzzy systems introduced in the paper are also highly 
relevant. They are the most common system used in applications and they
have close relation to neural networks \cite{Haykin08}. Duality in the representation 
of the radial rule bases leads to interesting synergies when conjunctive systems 
are syntactically transformed into the implicative ones \cite{Coufal2015-EANN15}.

\section{Preliminaries}
\label{prelim}
This section briefly reviews several notions from fuzzy set theory
in order to set up the employed notation. The review draws on textbooks
and handbooks on fuzzy set theory \citep{Wang97, Klir95, Handbook-RED,
Handbook-GREEN, Novak00} and other more or less specialized literature
\citep{Mesiar00, Hajek98}. In order to distinguish the fuzzy sets from 
the ordinary ones, we call the latter the \textit{crisp sets}.

\begin{definition}
Let $X$ be a crisp set called the universum. A fuzzy set $A$ specified on the 
universum $X$ is any mapping $\mu_A$ from $X$ to the interval $\mr{[0,1]}$.
This mapping is commonly called the membership function of the fuzzy set~$A$.
We use $A(x)$ instead of $\mu_A(x)$ to simplify the notation.
\end{definition}

\begin{definition}
$\F(X)$ stands for the set of all fuzzy sets specified on $X$.
\end{definition}

The special case of universa are the Cartesian products. Fuzzy sets specified
on the Cartesian products of crisp sets are called \textit{fuzzy relations}.

In the definitions below, we assume that the fuzzy set $A$ is specified on
some universe $X$.

\begin{definition}
The fuzzy set $A$ is called normal if there exists at least one element of $X$ such
that $A(x)=1$.
\end{definition}

\begin{definition}
The height of the fuzzy set $A$, denoted $\mr{height}(A)$, is the largest
membership degree to $A$, formally,
$\mr{height}(A)=\sup_{x\,\in X}\{A(x)\}$.
\end{definition}

\begin{definition}
The support of the fuzzy set $A$, denoted $\mr{supp}(A)$, is the crisp subset
of $X$ such that its elements have a non-zero membership degree to $A$, formally,
$\mr{supp}(A)=\{x\in X\,|\,A(x)>0\}$.
\end{definition}

\begin{definition}
The core of the fuzzy set $A$, denoted $\mr{core}(A)$, is the crisp subset
of the elements of $X$ such that $A(x)=1$,
formally, $\mr{core}(A)=\{x\in X\,|\,A(x)=1\}$.
\label{coreDef}
\end{definition}

\begin{definition}
Let $\alpha\in [0,1]$. The $\alpha$-cut of the fuzzy set $A$, denoted 
$[A]^\alpha$, is the crisp subset of $X$ specified as 
$[A]^{\alpha}=\{x\in X\,|\,A(x)\geq\alpha\}$.
\label{alphaDef}
\end{definition}

By inspecting the above definitions, one may identify several relations such
as $[A]^0=X$ or $[A]^1=\mr{core}(A)$.

\subsection{$t$-norms and fuzzy conjunctions}
In fuzzy set theory, operations on the fuzzy sets are tightly related to the 
notions of the $t$-norm, $s$-norm and fuzzy implication. We recall
their definitions and basic properties.

\begin{definition}
The $t$-norm $T$ is a function from  $[0,1]\times[0,1]$ to $[0,1]$
satisfying the following four properties\tn{:}\\[0.25cm]
\begin{tabular}{l}
\hspace*{0.5cm} $\mr{(T1)}$ $T(a,b)=T(b,a)$ (\textit{commutativity}),\\
\hspace*{0.5cm} $\mr{(T2)}$ $T(a,T(b,c))=
T(T(a,b),c)$ (\textit{associativity}),\\
\hspace*{0.5cm} $\mr{(T3)}$ \tn{if} $a_1 \leq a_2$ 
\tn{then} $T(a_1,b) \leq T(a_2,b)$ (\textit{monotonicity}),\\
\hspace*{0.5cm} $\mr{(T4)}$ $T(a,1)=a$ (\textit{boundary condition}).\\
\end{tabular}
\end{definition}

It is well known that the $t$-norms extend the classical Boolean conjunction
because one has $T(1,1)=1$ and $T(1,0)=T(0,1)=T(0,0)=0$. That is why
they are called the \textit{fuzzy conjunctions} and employed to specify
the fuzzy intersections of fuzzy sets.

\begin{definition}
\label{Fint}
Let $A_1,\dots A_m$, $m\in\N$ be the fuzzy sets specified on a common 
universum $X$. Their fuzzy intersection is the fuzzy set $\bigwedge_{j=1}^m A_j$
with the membership function specified on the basis of some $t$-norm $T$ as
\begin{equation}
(\bigwedge_{j=1}^m A_j)(x)=T(A_1(x),\dots,A_m(x)).
\end{equation}
\end{definition}

In the following sections, we will usually denote the $t$-norms by the star
symbol~$\star$ and use the multiplicative notation $T(a,b)=a\star b$.
Using this notation the aforementioned formula writes
\begin{equation}
(\bigwedge_{j=1}^m A_j)(x)=A_1(x)\star\,\dots\,\star A_m(x).
\end{equation}

Different $t$-norms determine different fuzzy conjunctions/interesections.
The best known examples of the $t$-norms are the minimum $t$-norm 
$T_{\mr{M}}(a,b)=\min\{a,b\}$, the product $t$-norm $T_{\mr{P}}(a,b)=a\cdot b$ 
and the {\L}ukasiewicz $t$-norm $T_{\mr{L}}(a,b)=\max\{0,a+b-1\}$
for $a,b\in[0,1]$.

Another often imposed requirement on the $t$-norms is their continuity:
(T5) $T$ is a continuous function.

The \textit{Archimedean \mb{$t$-norms}} play an important role in the theory of $t$-norms
\cite{Mesiar00, Klir95}. These $t$-norms possess the Archimedean property that states 
that for any $a,b\in(0,1)$ there exists an $n\in\N$ such that  (using the multiplicative
notation) $T(a,\dots,a)=a\star\dots\star a<b$, and $a$ is presented $n$-times in the
formula.

The continuous Archimedean $t$-norms form an important class because
of their well known characterization theorem \cite{Mesiar00}. In~order to review the 
theorem, we recall the concept of the continuous additive (decreasing) generator
of a $t$-norm.

\begin{definition}
\label{addgendef}
The $t$-norm $T$ has a continuous additive generator if there exists a
continuous, strictly decreasing function $t:[0,1]\rightarrow[0,+\infty]$,
$t(1)=0$, which is uniquely determined up to a positive
multiplicative constant, such that for all $a,b\in[0,1]$
\begin{equation}
T(a,b)=t^{(-1)}(t(a)+t(b))
\end{equation}
where $t^{(-1)}$ is the pseudo-inverse of $\;t$.
\end{definition}

\vspace{1ex}
The pseudo-inverse $t^{(-1)}$ of the continuous additive generator $t$
is the function from $[0,+\infty]$ to $[0,1]$ which is specified as
\begin{equation}
\label{psinvdef}
t^{(-1)}(z)=\left\{
\begin{array}{ll}
t^{-1}(z)&\;\;\;\tn{for}\;\;\;\;\;z\in[0,t(0)]\\
0&\;\;\;\tn{for}\;\;\;\;\;z\in(t(0),+\infty]\\
\end{array}
\right .
\end{equation}
where $t^{-1}$ is the ordinary inverse of $\;t$. 

Based on the above formula, one may specify the pseudo-inverse 
of the continuous additive generator using the ordinary inverse as
\begin{equation}
t^{(-1)}(z)=t^{-1}(\min\{t(0),z\}),\;\;\;z\in[0,\infty].
\end{equation}

Having the concept of the continuous additive generator defined, 
the representation theorem for the continuous Archimedean $t$-norms
reads as follows:

\begin{theorem}
T is a continuous Archimedean $t$-norm if and only if T has a~continuous
additive generator.
\end{theorem}

\noindent
\textbf{Proof.} See \cite{Mesiar00}, Sec. 5.1 on page 122.\hfill$\Box$

\subsection{$s$-norms and fuzzy disjunctions}
The counterparts to the $t$-norms are the so-called $s$-norms (or $t$-conorms). 
They extend the Boolean disjunction and are also called as the \textit{fuzzy disjunctions}.
A~similar definition to Definition~\ref{Fint} is established that defines the
\mbox{$s$-norms}, see~\cite{Klir95}. Because we will not work extensively with
the $s$-norms in this paper, we do not go into more details here. We only mention
that the $s$-norms are employed to establish the fuzzy unions of fuzzy sets:

\begin{definition}
\label{Funion}
Let $A_1,\dots A_m$, $m\in\N$ be fuzzy sets defined on a common universum $X$.
Their fuzzy union is the fuzzy set $\bigvee_{j=1}^m A_j$ with the membership function 
specified on the basis of some $s$-norm $S$ as
\begin{equation}
(\bigvee_{j=1}^m A_j)(x)=S(A_1(x),\dots,A_m(x)).
\end{equation}
\end{definition}
The most common $s$-norm used in applications is the maximum
$s$-norm, i.e., $S(a,b)=\max\{a,b\}$ for $a,b\in[0,1]$.

\subsection{Resiuated fuzzy implications}
In fuzzy set theory, fuzzy implications form several classes. They~are operations
from the unit square to the unit interval and extend the classical Boolean implication.
In this paper, we are interested in the so-called \textit{residuated implications}
or $R$-implications in short. The $R$-implications are derived from the
$t$-norms by the operation of residuation~\cite{Klir95, Hajek98}.

\begin{definition}
Let $\star$ be the $t$-norm. The residuated implication $\ra_{\star}$ is
the operation on the unit square specified as
\begin{equation}
a\ra_{\star}b=\sup\nolimits_z\{z\in[0,1]\,|z\star a\leq b\}.
\label{resDef}
\end{equation}
\end{definition}

Based on this definition, one can easily see that for any $R$-implication 
it holds  that $a\ra_{\star}b=1$ iff $a\leq b$.

The residuated implications that are derived from the minimum, product and
{\L}ukasiewicz \mb{$t$-norms} write for $a,b\in [0,1]$ as follows:
\begin{itemize}
\item
$a\ra_{\mr{M}}b= 1$ for $a\leq b$ and $a\ra_{\mr{M}}b=b$ for $a>b$ - 
the so-called G\"{o}del implication;
\item
$a\ra_{\mr{P}}b= 1$ for $a\leq b$ and $a\ra_{\mr{P}}b=b/a$ for $a>b$ - 
the so-called Goguen implication;
\item
$a\ra_{\mr{L}}b= 1$ for $a\leq b$ and $a\ra_{\mr{L}}b=1-a+b$ for $a>b$ -
the so-called {\L}ukasiewicz implication.
\end{itemize}

\section{Basics of fuzzy systems}
\label{Sec3}
In this section, we review the basics of fuzzy systems. More profound
reviews are available in \citep{Handbook-GREEN, Wang97, Yager94, Klir95}.
We introduce the general concept and specialize it to what is called 
the \textit{conjunctive and implicative fuzzy systems in the standard configuration}.
We present the computational models of both these systems.

\subsection{General fuzzy system}
The fuzzy system encodes a function from an input crisp set $X$ to an output
crisp set $Y$. A concrete computational form of this function draws on the mathematical
representations of the four building blocks of the fuzzy system: \textit{a~fuzzifier},
\textit{a~rule base}, \textit{an~inference engine} and \textit{a~defuzzifier}.
In the quadruple, the second and the third blocks constitute the so-called
\textit{proper fuzzy system}. It processes a fuzzy set on its input and yields a fuzzy set
on its output. The fuzzifier is just a device for mapping points of the input set $X$
to fuzzy sets specified on $X$. The defuzzifier then maps fuzzy sets specified on $Y$
into points of $Y$.

The rule base of the fuzzy system bears the knowledge stored in the system. 
Canonically, this knowledge is represented by a (crisp) set of IF-THEN rules. 
The IF-THEN rules represent fuzzy relations on the input-output set \mbox{$X\times Y$}.
These individual fuzzy relations are combined into the overall fuzzy relation $RB$
that mathematically represents the rule base. 

When the fuzzy system is in use, an input into the system $x^*\in X$ is fuzzified
in the fuzzifier to a fuzzy set $A'$ specified on $X$. The fuzzy set~$A'$ is 
combined with the stored knowledge in the inference engine. As the output, the
fuzzy set $B'\in\F(Y)$ is released. Thus, the inference engine performs a mapping 
from $\F(X)\times \F(X\times Y)$ to $\F(Y)$. Finally, the fuzzy set~$B'$ is defuzzified
in the defuzzifier into a point $y^*\in Y$ that constitutes the output from the
fuzzy system. This flow of computation determines the function computed by
the fuzzy system. Let us now state the formal definition of the fuzzy system.
\begin{definition}
The fuzzy system is the 6-tuple $\{X,Y,fuzz,RB,IE,defuzz\}$ where
\begin{itemize}
\item $X$ is an input crisp set,
\item $Y$ is an ouput crisp set,
\item $fuzz:X\ra \F(X)$ is a mapping from $X$ to $\F(X)$ (the fuzzifier),
\item $defuzz:\F(Y)\ra Y$ is a mapping from $\F(Y)$ to $Y$ (the defuzzifier),
\item the rule base $RB$ is a fuzzy relation specified on the Cartesian product
$X\times Y$, i.e., $RB\in\F(X\times Y)$,
\item
the inference engine IE is a mapping from
$\F(X)\times F(X\times Y)$ to $\F(Y)$.
\end{itemize}
The fuzzy system determines the function $FS:X\ra Y$ that is specified as
\begin{equation}
FS:y^*=defuzz(IE(fuzz(x^*),RB)).
\label{FSfunct}
\end{equation}
\label{FSdef}
\end{definition}

The above definition associates the fuzzy system $FS$ with a certain function.
If needed, this function will be explicitly indicated by using the argument
$x^*\in X$. That is, by $FS(x^*)$ we will mean the function~(\ref{FSfunct})
that is related to the fuzzy system $FS$.

In Definition~\ref{FSdef}, the rule base is specified on a general level as 
a fuzzy relation. In what follows, we will specialize this specification to the
notion of the MISO rule base.

\subsection{Conjunctive and implicative MISO rule bases}
\label{ConImplSec}
In Definion~\ref{FSdef}, we work with general input and output sets.
In real applications, however, the MISO (multiple-input single-output) 
configuration of the fuzzy system and its rule base is the most common.
Under this configuration, $X\subseteq\R^n$, $n\in\N$ and $Y\subseteq\R$;
and the corresponding rule bases are called the MISO rule bases.

There are two basic ways how individual IF-THEN rules are interpreted
and combined into the final rule base: the \textit{conjunctive} and the \textit{implicative}
way.

In the \textit{conjunctive rule bases}, the antecedents (the IF parts) and consequents 
(the THEN parts) of the IF-THEN rules are combined by a fuzzy conjunction and
individual rules by a fuzzy disjunction. The stored knowledge is viewed as the parallel list
(the fuzzy disjunction) of input-output examples (fuzzy points) from the relation
the rule base represents. The conjunctive rule bases are seen as driven by examples.

In the \textit{implicative rule bases}, the antecedents are combined with the 
consequents by a fuzzy implication and individual rules by a fuzzy disjunction.
In this case, the IF-THEN rules correspond to the set of restrictions the represented
relation must satisfy simultaneously (the fuzzy conjunction). This way of the
representation is seen as logically driven.

Let us be more specific. Let the rule base of the fuzzy system consist of 
$m\in\N$ IF-THEN rules. The $j$-th IF-THEN rule encodes the fuzzy relation~$R_j$
specified on the $X\times Y\subseteq\R^{n+1}$ set with the membership function
\begin{equation}
R_j(\bs{x},y)=A_{j1}(x_1)\star\dots\star A_{jn}(x_n)\rhd B_j(y).
\label{RuleLong}
\end{equation}
In the formula, $A_{ji}$, $j=1,\dots,m$, $i=1,\dots,n$ stand for the one-dimensional
fuzzy sets implementing the antecedent of the $j$-th rule and $B_j$ is the fuzzy set 
that implements the consequent of the rule. The $\star$ operation corresponds to
the fuzzy conjuction which is used to form the antecedent of the $j$-th rule
\begin{equation}
A_j(\bs{x})=A_{j1}(x_1)\star\dots\star A_{jn}(x_n).
\label{RuleAnt}
\end{equation}

The $\rhd$ operation corresponds either to the fuzzy conjunction $\star$ used in
(\ref{RuleAnt}) - for the conjunctive systems; or to the fuzzy implication $\ra_{\star}$ in
the case of the implicative systems. In the latter case, we stress that $\ra_{\star}$ is the
$R$-implication that is derived from the $t$-norm $\star$ that is used to build
the antecedents (\ref{RuleAnt}) for $j=1,\dots,m$.

The fuzzy relation (\ref{RuleLong}) can be expressed in the more compact form
emploing only the antecedent and consequent of the rule: 
\begin{equation}
R_j(\bs{x},y)=A_{j}(\bs{x})\rhd B_j(y).
\label{RuleShort}
\end{equation}
In the above formulas, clearly, one has $\bs{x}=(x_1,\dots,x_n)\in\R^n$ and $A_j$
is the multidimensional fuzzy set that is specified on the input set $X\subseteq\R^n$.

Using the notation of (\ref{RuleShort}), one has the following two formulas to represent
the conjunctive and implicative MISO rule bases:
$$
RB_{conj}(\bs{x},y)=\bigvee_{j=1}^m\{ A_j(\bs{x})\star B_j(y)\},\;\;\;
RB_{impl}(\bs{x},y)=\bigwedge_{j=1}^m\{ A_j(\bs{x})\ra B_j(y)\}.
$$

In applications, the maximum $s$-norm is usually taken to interpret the
fuzzy disjunction $\bigvee$; and the minimum $t$-norm to interpret the
fuzzy intersection $\bigwedge$. This leads to the following formal definition.

\begin{definition}
The conjunctive or the implicative MISO rule base is the 8-tuple
$RB_{\rhd}=\{X,Y,n,m,\,\star\,,\{\{A_{ji}\}_{i=1}^n\}_{j=1}^m,
\{B_j\}_{j=1}^m,\rhd\}$
where
\begin{itemize}
\item $X\subseteq\R^n$ is an input crisp set,
\item $Y\subseteq\R$ is an output crisp set,
\item $n\in\N$ is the dimension of $X$,
\item $m\in\N$ is the number of \tn{IF-THEN} rules consituting the rule base,
\item $\star$ is a $t$-norm,
\item $\{\{A_{ji}\}_{i=1}^n\}_{j=1}^m$ is the set of $n\cdot m$ one-dimensional fuzzy sets 
forming the antecedents of \tn{IF-THEN} rules where $A_j=A_{j1}\star\dots\star A_{jn}$,
$j=1,\dots,m$, i.e., $A_j\in\F(X\subseteq\R^n)$, $j=1,\dots,m$,
\item $\{B_j\}_{j=1}^m$ is the set of $m$ fuzzy sets forming the consequents
of \tn{IF-THEN} rules, i.e., $B_j\in\F(Y\subseteq\R)$, $j=1,\dots,m$,
\item $\rhd$ corresponds either to the $t$-norm $\star$, i.e., $\rhd=\star$,
for the conjunctive rule base; or to the residuated fuzzy implication derived 
from $\star$, i.e., $\rhd=\,\ra_{\star}$,
for the implicative rule base, respectively.
\end{itemize}
The specifications of the corresponding fuzzy relations write as
\begin{equation}
RB_{conj}(\bs{x},y)=\max\nolimits_{j=1}^m\{ A_j(\bs{x})\star B_j(y)\}
\end{equation}
for the conjunctive rule base; and
\begin{equation}
RB_{impl}(\bs{x},y)=\min\nolimits_{j=1}^m\{ A_j(\bs{x})\ra B_j(y)\}
\end{equation}
for the implicative rule base.
\end{definition}

By inspecting the specification of the $\rhd$ operation in the above definition, we see that 
in the conjunctive rule bases, an individual rule is build up using the identical $t$-norm 
$\star$. That is, the identical $\star$ is used to build up antecedents and to combine them
with the consequents. In the implicative rule bases, the operation of residuation
(\ref{resDef}) gives the relation between the $\star$ of the antecedent (\ref{RuleAnt})
and the employed residuated fuzzy implication $\ra_{\star}$.

\subsection{Singleton fuzzifier and CRI engine}
We proceed our specialization by recalling the \textit{singleton fuzzifier} and 
the \textit{CRI engine} that are the most common choices for the fuzzifier and
the inference engine in applications.
\begin{definition}
Let $x^*\in X$ be the input to the fuzzy system. The singleton fuzzifier relates
its input to the fuzzy set $A'$ with the membership function
\begin{equation}
A'(x)=
\left\{
\begin{tabular}{cc}
$1$&$\mathrm{for}$ $x=x^*$\\
$0$&$\mathrm{otherwise.}$
\end{tabular}
\right.
\label{Sfuzz}
\end{equation}
\label{SfuzzDef}
\end{definition}

The CRI engine implements the well known \textit{compositional rule of inference}
\cite{Wang97, Klir95} for composing a fuzzy set with a fuzzy relation. The definition
of the engine reads as follows:
\begin{definition}
Let $A$ be a fuzzy set specified on $X$ and $RB$ a fuzzy relation specified on $X\times Y$.
The CRI engine specifies the membership function of the output fuzzy set $B'$, $B'\in\F(Y)$, as
\begin{equation}
B'(y)=\sup_{x\in X}\{A(x)\star RB_{\rhd}(x,y)\}
\label{CRIdef}
\end{equation}
for $A\in\F(X)$, $RB\in\F(X\times Y)$ and $\star$ being some $t$-norm.
\end{definition}

The popularity of the combination of the singleton fuzzifier and the CRI engine
stems from the fact that when $A'$ is specified according to (\ref{Sfuzz}), then
(\ref{CRIdef}) writes as
\begin{equation}
B'(y)=RB(x^*,y).
\end{equation}
This is due to the properties of the $t$-norms, namely that $T(0,a)=0$ and $T(1,a)=a$
for any $t$-norm $T$ and $a\in[0,1]$. Clearly, this choice greatly simplifies the computation
of the CRI engine; and it does not depend on the explicit form of $\star$ in (\ref{CRIdef}).
However, there are other inference engines available. As an example, let us mention
the engine that is based on the Bandler-Kohout subproduct \citep{BKohout2010}.

\subsection{WA and MOM methods of defuzzification}
Discussing methods of defuzzification, there is plenty of them proposed
in literature \cite{Wang97, Klir95, Driankov93}. Here we mention the two 
that are suitable for the MISO systems where $Y\subseteq\R$. The first
is the \textit{weighted average method} (the WA method) which is the simplified
version of the COG defuzzification method. The method has several other
names in literature such as the center average defuzzifier~\cite{Wang97}
or the method of heights \cite{Driankov93}. We later relate the WA method
to the conjunctive fuzzy systems. The second method of defuzzification,
which we later relate to the implicative fuzzy systems, is 
the \textit{mean-of-maxima method} (the MOM method).

The center of gravity (COG) method of defuzzification states the defuzzified
value $y^*_{B}$ of the $B\in\F(Y)$ fuzzy set
as $y^*_{B}=\int_Y y\cdot B(y)\, dy/\int_Y B(y)\,dy$. The point $y^*_{B}$
is known as the \textit{centroid} of the $B$ fuzzy set. The COG method is generally
computationally intensive and it might be hard to state $y^*_{B}$ explicitly.

To ease the computation, one may take the advantage  of the singleton fuzzifier. 
In this case and for the conjunctive systems, one has
$$
RB_{conj}(x^*,y)=\max\nolimits_{j=1}^m \{A_j(x^*)\star B_j(y)\}´
=\max\nolimits_{j=1}^m \{B'_j(y)\}
$$
where we denoted $B'_j(y)=A_j(x^*)\star B_j(y)$.
A simplification to the COG method is done by considering the weighted average of 
the centroids of the consequent fuzzy sets $B_j$. These centroids can
be computed in advance and do not change with the input into the fuzzy system 
because the input affects only the antecedents of rules $A_j(x^*)$. The weights
in the average are determined as the heights of the $B'_j$ fuzzy sets. 
If the consequent fuzzy sets $B_j$ are normal, one has $\tn{height}(B'_j)=A_j(x^*)$
for each $j=1,\dots,m$. Following this line of thinking carries out the formal definition
of the WA method of defuzzification.

\begin{definition}
Let $Y\subseteq\R$ and $B(y)=\max\nolimits_{j=1}^m \{B'_j(y)\}$ where
$B'_j(y)=A_j(x^*)\star B_j(y)$ for a given input $x^*\in X$. Let there exist 
at least one $j\in\{1,\dots,m\}$ such that $A_j(x^*)>0$. Let the centroids 
of $B_j$ sets, $y^*_{B_j}=\int y\cdot B_j(y)\,dy/\int B_j(y)\, dy$, exist and
$y^*_{B_j}\in\R$ for all $j=1,\dots,m$.
Then the weighted average method of defuzzification is specified as
\begin{equation}
y^*=\frac{\sum_{j=1}^m A_j(x^*)\cdot y^*_{B_j}}
{\sum_{j=1}^m A_j(x^*)}\, .
\label{CMdefuzz}
\end{equation}
\end{definition}
By inspecting the above formula, we see that it is defined only when the centroids
$y^*_{B_j}$ exist and for those inputs for which there exists at least one
rule $j$ that is firing ($A_j(x^*)>0$); because otherwise we would have 
the denominator equal to zero in~(\ref{CMdefuzz}).

The denominator is zero if none of the rules is firing, i.e., $A_j(x^*)=0$ for
all $j=1,\dots,m$. What should be the output of the fuzzy system in this situation?
The technical solution could be considering the limit of (\ref{CMdefuzz}) for
the denominator approaching zero, but it does not exist. However, considering
the problem methodologically, we should not allow that this situation happens,
because if it happened, then it would mean that the knowledge stored in
the fuzzy system is irrelevant to such the input. Adopting the idea that the stored
knowledge must be relevant to all possible inputs leads to the requirement that
for every $x^*\in X$ there exists at least one rule $j$ such that $A_j(x^*)>0$.
Such rule bases are called complete \cite{Wang97, Driankov93}.

\begin{definition}
Degree of covering of a rule base is specified as
\begin{equation}
\mr{DOC}=\inf_{x\in X}\{\max\nolimits_{j=1}^m\{A_j(x)\}\}.
\label{DOC}
\end{equation}
The rule base of a fuzzy system is called complete if $\mr{DOC}>0$.
\label{DOCdef}
\end{definition}
Note that the notions of DOC and completeness are relevant to the general
rule bases, not only to the MISO ones. Now, we are going to switch to
the MOM method of defuzzification.

\begin{definition}
Let $Y\subseteq\R$ and $B\in\F(Y)$ be normal. Let both $\inf\{\tn{core}(B)\}$
and $\sup\{\tn{core}(B)\}$ be finite. Then the mean-of-maxima defuzzifier 
is specified~as
\begin{equation}
y^*=\frac{\inf\{\tn{core}(B)\}+\sup\{\tn{core}(B)\}}{2}.
\label{MOMdefuzz}
\end{equation}
\end{definition}

In the above definition, we assume that both infimum and supremum of $\tn{core}(B)$
are finite. Later we will see that in the case of the implicative fuzzy systems, the $\tn{core}(B)$
is specified as the intersection of the cores of the fuzzy sets that are derived from the
consequents $B_j$; and the requirement on finiteness can be assured by a reasonable
choice of the $B_j$ sets. For example, we may require that they form the fuzzy numbers
\cite{Mesiar00, Klir95, Wang97}.

\subsection{Conjunctive and implicative fuzzy systems in the standard configuration}
\label{FSstd}
We end the section by formally introducing the \textit{conjunctive and implicative
fuzzy system in the standard configuration}. We call them in the standard configuration
because this configuration prevails in the real-world applications.

\begin{definition}
Let FS be the fuzzy system. We call it the conjunctive fuzzy system in the standard 
configuration if $X\subseteq\R^n$, $Y\subseteq\R$. FS has the conjunctive MISO
rule base, uses the CRI inference engine and employs the singleton fuzzifier and
the WA method of defuzzification.
\label{FSconjSTD}
\end{definition}

\begin{definition}
Let FS be the fuzzy system. We call it the implicative fuzzy system in the standard 
configuration if $X\subseteq\R^n$, $Y\subseteq\R$. FS has the implicative MISO
rule base, uses the CRI inference engine and employs the singleton fuzzifier and 
the MOM method of defuzzification.
\label{FSimplSTD}
\end{definition}

In what follows, we are interested in the computational models of both types
of these fuzzy systems, i.e., in the explicit forms of the associated 
$FS:X\subseteq\R^n\ra Y\subseteq\R$ functions.

\begin{lemma}
Let $FS_{conj}$ be the conjunctive fuzzy system in the standard configuration. Let the rule
base of the fuzzy system be complete and the centroids $y^*_{B_j}\in\R$ of the $B_j$
sets exists for all $j=1,\dots,m$. Then the fuzzy system computes the function
\begin{equation}
FS_{conj}(\bs{x}^*)=\frac{\sum_{j=1}^m A_j(\bs{x}^*)\cdot y^*_{B_j}}
{\sum_{j=1}^m A_j(\bs{x}^*)}.
\label{FSconj}
\end{equation}
\label{LemmaC}
\end{lemma}

\noindent
\textbf{Proof.}
If the singleton fuzzifier is employed, then the output of the CRI inference
engine writes $B'(y)=RB(\bs{x}^*,y)=\max_{j=1}^m \{A_j(\bs{x}^*)\star B_j(y)\}$.
The application of the WA method of defuzzification (\ref{CMdefuzz}) gives the output 
(\ref{FSconj}) under two assumptions. The first one is the completeness of the rule base;
and the other is the existence and finiteness of the centroids of the $B_j$ sets.\hfill$\Box$

When we switch to the implicative fuzzy systems, we will find as crucial
the notion of \textit{coherence of the implicative fuzzy system} \cite{Dubois97, Driankov93}.
In words, coherence assures that for each possible input there is specified a fully 
reliable output from the implicative fuzzy system. Mathematically, we require that the 
output fuzzy set from the inference engine is always normal. This is equivalent to
the requirement that the core of the inference engine's output fuzzy set is non-empty,
as the emptiness would mean that for the given input we have contradictory
rules in the rule base.

\begin{definition}
\label{cohdef}
The implicative fuzzy system in the standard configuration is \tn{\textbf{coherent}}
if for any input $\bs{x}^*\in X$ the fuzzy set $B'$ that is released from the inference
engine is normal.
\end{definition}

The following two lemmas state equivalent conditions for coherence and bring
a certain insight into the computational model of the implicative fuzzy systems in
the standard configuration.

\begin{lemma}
For any input $\bs{x}^*\in X$ into the implicative fuzzy system in the standard
configuration, the core of the $B'$ fuzzy set writes as
\begin{equation}
\label{coreB}
\mr{core}(B')=\bigcap_{j=1}^m \;[B_j]^{A_j(\bs{x}^*)}
\end{equation}
where on the right-hand side there is the intersection of the crisp sets 
$[B_j]^{A_j(\bs{x}^*)}$ which are the $A_j(\bs{x}^*)$-cuts of the consequents
fuzzy sets $B_j$.
\label{coreBlemma}
\end{lemma}

\noindent
\textbf{Proof.} Because the fuzzy system is in the standard configuration, one has
$B'(y)=\min_{j=1}^m \{A_j(\bs{x}^*)\ra B_j(y)\}$. Furthermore, we have
$[B_j]^{A_j(\bs{x}^*)}=\{ y\,|\,A_j(\bs{x}^*)\leq B_j(y) \}$ by Definition~\ref{alphaDef}
of the $\alpha$-cut.

1) Let us show that $\mr{core}(B')\subseteq\bigcap_{j=1}^m \;[B_j]^{A_j(\bs{x}^*)}$.
If $\mr{core}(B')=\emptyset$, then it trivially holds. Let $y\in\mr{core}(B')$, i.e.,
$B'(y)=1$, then due to the specification of the $B'$ set, $B'(y)=\min_j\{A_j(\bs{x}^*)\ra B_j(y)\}$, 
and the properties of the $t$-norms ($T(a_1,\dots,a_n)=a_1\star\dots\star a_n=1$ iff 
$a_j=1$ for each~$j$), it must be  $B'_j(y)=1$ for every $j$ as well.
For the residuated implications one has
$I(a,b)=1$ iff $a\leq b$, thus the inequality $A_j(\bs{x}^*)\leq B_j(y)$ holds
for every~$j$. In other words, $y\in [B_j]^{A_j(\bs{x}^*)}$ for all $j=1,\dots,m$
and therefore $y\in \bigcap_{j=1}^m \;[B_j]^{A_j(\bs{x}^*)}$.

2) Let us show that $\bigcap_{j=1}^m \;[B_j]^{A_j(\bs{x}^*)}\subseteq\mr{core}(B')$.
If $\bigcap_{j=1}^m \;[B_j]^{A_j(\bs{x}^*)}=\emptyset$, then it trivially holds.
Let $y\in\bigcap_{j=1}^m \;[B_j]^{A_j(\bs{x}^*)}$, then $A_j(\bs{x}^*)\leq B_j(y)$
holds for every~$j=1,\dots, m$. So one has also $\min_{j=1}^m \{A_j(\bs{x}^*)\ra B(y)\}=1$ 
and therefore $y\in\tn{core}(B')$.\hfill$\Box$

\begin{theorem}
The implicative fuzzy system in the standard configuration is coherent if and only if 
for any $\bs{x}^*\in X$ it holds that
\begin{equation}
\label{cuts}
\bigcap_{j=1}^m \;[B_j]^{A_j(\bs{x}^*)}\not=\emptyset.
\end{equation}
\label{ctbasic}
\end{theorem}

\noindent
\textbf{Proof.} Obviously, a fuzzy set is normal if and only if its core is non-empty.
From the form of the core of $B'$ in the implicative fuzzy systems in the standard
configuration we have the statement of the theorem.\hfill$\Box$

Having the notion of coherence defined, we may approach to the specification
of the computational model of the implicative fuzzy systems in the standard
configuration.

\begin{lemma}
Let $FS_{impl}$ be the coherent implicative fuzzy system in the standard 
configuration and both $\inf\{\tn{core}(B')\}$ and $\sup\{\tn{core}(B')\}$
be finite for any input $\bs{x}^*\in X$. Then the fuzzy system computes
the function
\begin{equation}
FS_{impl}(\bs{x}^*)=\frac{\inf\{\bigcap_{j=1}^m [B_j]^{A_j(\bs{x}^*)}\}
+\sup\{\bigcap_{j=1}^m [B_j]^{A_j(\bs{x}^*)}\}}{2}.
\label{IFScomp}
\end{equation}
\label{lemma2}
\end{lemma}

\vspace{-1ex}
\noindent
\textbf{Proof.} 
Clearly, the formula (\ref{IFScomp}) is the result of when the assertion of 
Lemma~\ref{coreBlemma} is applied to the formula (\ref{MOMdefuzz}). Note that
we have $B'$ set always normal due to the assumed coherence of the system.\hfill$\Box$

From the above lemmas, we see that in order to have the concrete computational
models effectively specified one has to meet several assumptions. In the case 
of the conjunctive fuzzy systems, these are the assumptions of the completeness
of the rule base and specification of the finite centroids. 
In the case of the implicative fuzzy systems, the coherence is crucial because 
one wants to have consistent rules in the rule base. Further, both minima and
suprema of $\mr{core}(B')$ must be finite. In what follows we will show that 
these assumptions can be treated in a rather convenient way in the class
of the radial fuzzy systems.

\section{Radial fuzzy systems}
\label{Sec4}
Radial fuzzy systems form a subclass of the fuzzy systems in the standard
configuration. Hence, we have $X\subseteq\R^n$ and $Y\subseteq\R$ in these systems.
The fuzzy sets employed within the radial fuzzy systems are implemented using radial
functions. Moreover, in the radial fuzzy systems, antecedents of the IF-THEN rules
are required to exhibit a certain shape preservation property.

The specification of the radial fuzzy systems draws on the specification
of the radial rule bases. Before we state the associated definition, we recall
the concepts of the \textit{radial function} and the \textit{scaled $\ell_p$ norm}.

\subsection{Radial functions and scaled $\ell_p$ norms}
In words, the radial function is a function that is invariant to changes in its argument
that do not affect the argument's distance to a central point of the function.
A prototypical example of such a change is rotation. Mathematically, this property 
is encoded by incorporating a norm of a difference between the argument and 
the selected central point. The distance of the argument from the central point 
is further modified by another univariate function.

\begin{definition}
A function $f:\R^n\ra\R$, $n\in\N$ is called radial if it has the form 
$f(\bs{x})=\Phi(||\bs{x}-\bs{a}||)$ where $||\cdot||$ is a norm in 
$\R^n$, $\Phi$ is a function from $[0,+\infty)$ to $\R$ and 
$\bs{a}\in\R^n$ is the central point of the radial function $f$.
\label{RadFceDef}
\end{definition}

In the above definition, the function $\Phi$ is sometimes called 
the \textit{shape function} and is usually non-negative and monotonic.

Concerning norms in $\R^n$, the best known class is the class of the 
$\ell_p$ norms. We extend this class by
introducing the class of the scaled $\ell_p$ norms.

\begin{definition}
Let $\bs{b}=(b_1,\dots,b_n)$, $b_i> 0$, $i=1,\dots, n$, $n\in\N$. The scaled
$\ell_p$~norm $||\cdot||_{p,\bs{b}}$ is the norm specified for $p\in[1,\infty]$
as follows\tn{:}
\begin{equation}
\begin{array}{l}
||\bs{u}||_{p,\bs{b}}\;=\;(|u_1/b_1|^p+\dots+|u_n/b_n|^p)^{1/p}\;\;\;\;
\mr{for}\;\;\;\;p\in[1,+\infty),\\[0.1cm]
||\bs{u}||_{\infty,\bs{b}}\;=\;\lim_{p\rightarrow\infty} ||\bs{u}||_{p,\bs{b}}=
\max_i \{|u_i/b_i|\}.
\end{array}
\end{equation}
\end{definition}

As for the standard $\ell_p$ norms, the settings $p=1$, $p=2$ and $p=\infty$ 
provide the most frequently used selections. The corresponding norms are 
the scaled $\ell_1$ (or octaedric), scaled $\ell_2$ (or Euclidean) and scaled 
$\ell_{\infty}$ (or cubic norm), respectively. In the first and the third case, the 
names are derived from the shapes of the unit ball which is the octaeder
or cube in $\R^3$ space,~respectively.

\subsection{Radial rule bases}
Here we put forward the notion of the \textit{radial rule base} that is subsequently
used to introduce the radial fuzzy systems.

\begin{definition}
\label{raddef}
The MISO rule base
$RB_{\rhd}=\{X,Y,n,m,\,\star\,,\{\{A_{ji}\}_{i=1}^n\}_{j=1}^m,$
$\{B_j\}_{j=1}^m, \rhd\}$ is called radial if the following three conditions 
are satisfied\tn{:}\\[0.2cm]
\tn{(i)} There exists a continuous function 
$act:[0,+\infty)\rightarrow [0,1]$, $act(0)=1$ such that
\tn{(a)} either there exists $z_0\in(0,+\infty)$ such that $act$ is strictly
decreasing on $[0,z_0]$ and $act(z)=0$ for $z\in[z_0,+\infty)$ or
\tn{(b)} $act$ is strictly decreasing on $[0,+\infty)$ and
$\lim_{z\rightarrow+\infty} act(z)=0$.\\[0.2cm]
\tn{(ii)} Fuzzy sets in the antecedent and consequent of the $j$-th
rule are specified~as
\vspace*{-0.3cm}
\begin{eqnarray}
\label{RadASet}
A_{ji}(x_i)&=&act\left(\frac{|x_i-a_{ji}|}{b_{ji}}\right),\\
\label{RadBSet}
B_j(y)&=&act\left(\frac{\max\{0,|y-c_j|-s_j\}}{d_j}\right)
\vspace{0.5pt}
\end{eqnarray}
where $n,m\in\N$\tn{;} $i=1,\dots,n$\tn{;} $j=1,\dots,m$\tn{;}
$\bs{x}\in\R^n$, $\bs{x}=(x_1,\dots,x_n)$\tn{;}
$\bs{a}_j\in\R^n$, $\bs{a}_j=(a_{j1},\dots,a_{jn})$\tn{;}
$\bs{b}_j\in\R_+^n$, $\bs{b}_j=(b_{j1},\dots,b_{jn})$, $b_{ji}>0$\tn{;}
$c_j\in\R$\tn{;} $d_j\in\R$, $d_j>0$\tn{;} $s_j\in\R$, $s_j\geq 0$.\\[0.2cm]
\tn{(iii)} For each $\bs{x}\in X\subseteq\R^n$ the radial property holds, i.e.,
\begin{equation}
\label{RadProp}
A_j(\bs{x})=A_{j1}(x_1)\star\dots\star A_{jn}(x_n)=
act(\,||\bs{x}-\bs{a}_j||_{p,\bs{b}_j}),
\end{equation}
where $||\cdot||_{p,\bs{b}_j}$ is the scaled version of an $\ell_p$ norm $||\cdot||_p$.
This $\ell_p$ norm is common to all rules in the rule base.
\end{definition}

Generally speaking, in the radial fuzzy systems, the IF-THEN rules have specific
form that makes their computational model simpler and mathematically tractable,
especially in the case of the implicative systems as we will see later.

\begin{figure}[!htb]
\centering
\includegraphics[width=5.3in]{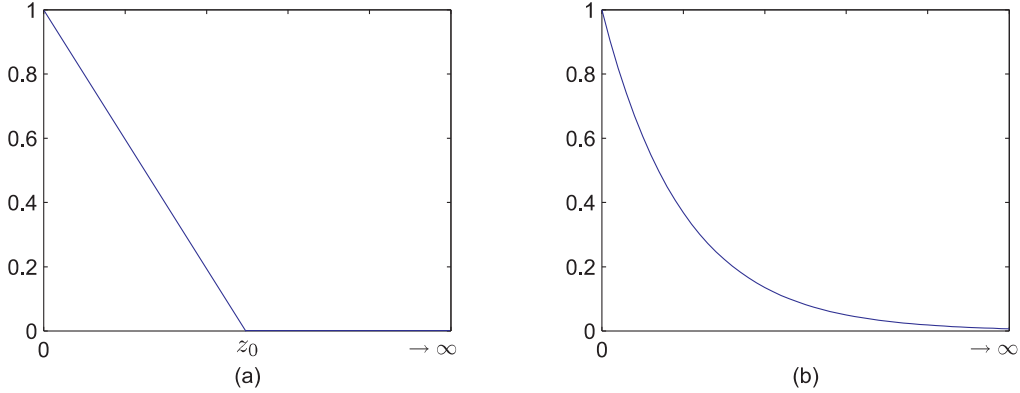}
\caption{Examples of the $act$ function.}
\label{fig_act}
\end{figure}

In the definition, there are presented three conditions that make a MISO rule base
to be radial. The first two conditions (i) and (ii) deal with the membership functions
of the fuzzy sets forming the antecedents and consequents of the IF-THEN rules.
Since the $act$ function is required to be a~non-increasing continuous function from 
$[0,+\infty)$ to $[0,1]$ such that $act(0)=1$, $\lim_{z\rightarrow+\infty} act(z)=0$,
see Fig.~\ref{fig_act}, the particular fuzzy sets form the fuzzy numbers
\cite{Mesiar00, Klir95, Wang97}.

In Definition~\ref{RadFceDef}, considering $n=1$, the
$\ell_1$ norm (the absolute value) and setting $\Phi(z)=act(z/b_{ji})$ 
or $\Phi(z)=act(\max\{0,z-s_j\}/d_j)$ for $z\in [0,+\infty)$ we get the specifications
of the one-dimensional fuzzy sets in condition (ii). Hence, in the radial rule bases,
the membership functions of one-dimensional fuzzy sets are the radial functions.

The difference between the specification of the antecedents and consequents
lies in the fact that the antecedents correspond to the generalized triangular
fuzzy numbers and the consequents to the generalized trapezoidal fuzzy numbers.
See graphical examples in Section~\ref{radSec}.

The third condition (iii), which is called the \textit{radial property}
and expressed by the equality (\ref{RadProp}), is crucial for
the specification of the radial fuzzy systems. 

\subsection{Radial property}
The radial property says that the combination of the one-dimensional fuzzy sets
preserves the radial shape. That is, by functional equation (\ref{RadProp}) it is 
required that the combination of the one-dimensional fuzzy sets $A_{ji}$ by the
employed $t$-norm~$\star$ must form a~multidimensional fuzzy set determined by 
the application of the $act$ function (the same as it is used for the $A_{ji}$ sets)
on the scaled $\ell_p$ norm \mbox{$||\cdot||_{p,\bs{b}_j}$} of the distance of 
the argument~$\bs{x}$ from the central point~$\bs{a}_j$.
Moreover, it is required that the scaling vector $\bs{b}_j$ is composed from
the parameters $b_{ji}$, $i=1,\dots, n$ of the individual one-dimensional
fuzzy sets, and the central point~$\bs{a}_j$ from the respective central
points~$a_{ji}$, $i=1,\dots,n$.

The radial property (\ref{RadProp}) is non-trivial because the selection of the
\mbox{$t$-norm~$\star$} and specification of the $A_{ji}$ sets already
determines the multidimensional membership function of the antecedent $A_j$. 

\textbf{Example.}
\textit{Non-triviality of the radial property.}
In this example, we will show that the combination of the triangular 
fuzzy sets by product does not exhibit the radial property.

Let the antecedent of the IF-THEN rule be formed by two triangular fuzzy~sets:
$$
A_1(x_1)=\max\{0,1-|x_1-a_1|/b_1\},\;\;\;
A_2(x_2)=\max\{0,1-|x_2-a_2|/b_2\};
$$
and let the employed $t$-norm $\star$ be the product $t$-norm.
Obviously, both sets satisfy conditions (i) and (ii) of Definition~\ref{raddef}
and the $act$ function corresponds to $act(z)=\max\{0,1-z\}$, $z\in[0,+\infty)$.
Using the product $t$-norm, the membership function of
the antecedent writes
$$
A(\bs{x})=\max\{0,1-|x_1-a_1|/b_1\}\cdot
\max\{0,1-|x_2-a_2|/b_2\}.
$$
Let us show that there is no $\ell_p$ norm such that 
(\ref{RadProp}) holds for every $\bs{x}\in\R^2$.

Denoting $u_1=(x_1-a_1)/b_1$, $u_2=(x_2-a_2)/b_2$,
the radial property (\ref{RadProp}) reads as
\begin{equation}
\label{cr01}
\max\{0,1-|u_1|\}\cdot
\max\{0,1-|u_2|\}=
\max\{0,1-||(u_1,u_2)||_p\}.
\end{equation}
Let $\bs{x}=(x_1,x_2)$ be set in such a way that $u_1=u_2=u\in(0,1)$
and assume that (\ref{cr01}) holds. We get
\begin{equation}
\label{uueq}
(1-|u|)^2=1-||(u,u)||_p.
\end{equation}
Note that the right-hand side cannot be zero because the left-hand side is 
positive for our choice of $u\in(0,1)$. The equality (\ref{uueq}) makes the value 
of the norm $||(u,u)||_p$ to be $||(u,u)||_p=1-(1-|u|)^2$; and
therefore
\begin{equation}
||(u,u)||_p=2|u|-|u|^2.
\label{su}
\end{equation}

It is clear that (\ref{su}) cannot specify a norm because
for any $u\in(0,1)$ and $\lambda\in(0,1)$ we obtain
\begin{equation}
||(\lambda u,\lambda u)\,||_p=2|\lambda u|-|\lambda u|^2
=|\lambda|\cdot(2|u|-|\lambda||u|^2),\\
\end{equation}
which contradicts the homogeneity property of the norms that requires that
$||c\bs{u}||=|c|\cdot ||\bs{u}||$ for any $c\in\R$.\\

By this example we see that the radial property is non-trivial and we
na-turally ask which the $t$-norms and shapes (the $act$ functions) can be
combined so that the radial property holds? This is the topic of the 
next section.

To end this section, remark that if the radial property holds, then the antecedents 
$A_j$ are represented by the multivariate radial functions. Indeed, if 
the radial property holds, then we have
$A_j(\bs{x})=act(||\bs{x}-\bs{a}_j||_{p,\bs{b}_j})$,
which fits the specification of the radial function given in Definition~\ref{RadFceDef}.
A link to the radial basis functions of RBF neural networks theory can be established
here \cite{Haykin08}.

Moreover, the radial property brings other advantages. Let us mention the following two.
First, the computational model of the related radial fuzzy systems, especially of the
implicative ones, is mathematically tractable. Second, when creating a rule base 
from data, the clustering techniques are commonly used to do the job, e.g., 
the fuzzy \mbox{$c$-means} algorithm is of popular use in this area \cite{Hoppner97}. 
The identified clusters then correspond to the antecedents and consequents of the
IF-THEN rules and they can be straightforwardly transformed into the membership
functions of the radial fuzzy sets.

\subsection{Building radial rule bases}
Having the concept of the radial rule base specified, there is a natural request
for examples of such the rule bases, especially with respect to the some previously
selected $t$-norms. Let us start with the minimum $t$-norm.

\begin{theorem}
\label{RadTheorem1}
Let the MISO rule base with $m$ rules be specified as follows\tn{:}\\[2pt]
\tn{1)} The employed $\;t$-norm is the minimum $t$-norm.\\[2pt]
\tn{2)} The $act$ function is any function from $[0,+\infty)$ to $[0,1]$
meeting the condition \tn{(i)} of Definition \tn{\ref{raddef}}.\\[2pt]
\tn{3)} The fuzzy sets forming the antecedents and consequents have the form
\begin{equation}
\label{FSdef1}
A_{ji}(x_i)=act\left(\frac{|x_i-a_{ji}|}{b_{ji}}\right),
\;\;\;\;
B_j(y)=act\left(\frac{\max\{0,|y-c_j|-s_j\}}{d_j}\right)
\end{equation}
for $j=1,\dots, m$\tn{;} $\bs{a}_j\in\R^n$\tn{;} $\bs{b}_j\in\R^n_+$\tn{;} 
$c_j\in\R$\tn{;} $d_j\in\R$, $d_j>0$\tn{;} $s_j\in\R$, $s_j\geq 0$.\\[0.2cm]
Then the rule base is radial.
\end{theorem}

\noindent
\textbf{Proof.}
The conditions 2) and 3) of the specification coincide with the conditions (i) and (ii)
of Definition \ref{raddef}. Hence it remains to prove that using the minimum
$t$-norm validates the radial property (\ref{RadProp}).

As the $act$ function is non-increasing we have 
$$
\min\{act(u_1),\dots,act(u_n)\}=act(\max\{u_1,\dots,u_n\})
$$
for $u_i\in[0,+\infty)$, $i=1,\dots,n$.
Therefore, if the conditions 2) and 3) are satisfied and we use
the minimum $t$-norm we obtain the antecedents in form
\begin{equation*}
A_j(\bs{x})=\min_i \left\{ act\left(\frac{|x_i-a_{ji}|}{b_{ji}}\right) \right\}=
act \left( \max_i\left\{\frac{|x_i-a_{ji}|}{b_{ji}}\right\} \right).
\end{equation*}
Because $b_{ji}>0$ for every $j$ and $i$, we can use the scaled cubic norm
to rewrite the above equality as
\begin{equation*}
A_j(\bs{x})=\min_i \left\{ act\left(\frac{|x_i-a_{ji}|}{b_{ji}}\right) \right\}=
act(\,||\bs{x}-\bs{a}_j||_{\infty,\bs{b}_j}),
\end{equation*}
which is exactly the radial property (iii) of Definition~\ref{raddef}.\hfill$\Box$

We see that for the minimum $t$-norm we can assure the validity of the radial 
property by a suitable, but non-restrictive choice of the $act$ function. 
Now, we aim at the continuous Archimedean $t$-norms.

\begin{theorem}
\label{RadTheorem2}
Let the MISO rule base with $m$ rules be specified as follows\tn{:}\\[2pt]
\tn{1)} The employed $\,t$-norm $\,T$ is continuous Archimedean with 
the continuous additive generator $\,t$
and its pseudo-inverse $t^{(-1)}$.\\[2pt]
\tn{2)} For parameters $q>0$ and $p\in[1,+\infty)$ the $act$ function 
has the form
\begin{equation}
\label{actArch}
act(z)=t^{(-1)}(qz^p),\;\;\;z\in[0,+\infty).
\end{equation}
\tn{3)} The fuzzy sets forming the antecedents and consequents are specified as
\begin{equation}
A_{ji}(x_i)=t^{(-1)}\!\!\left[q\!\left(\frac{|x_i-a_{ji}|}{b_{ji}}\!\right)^{\!\!p}\;\right]\!,
B_j(y)=t^{(-1)}\!\!\left[q\!\left(\frac{\max\{0,|y-c_j|-s_j\}}{d_j}\!\right)^{\!\!p}\;\right]\!.
\label{ArchDef}
\end{equation}
for $j=1,\dots, m$\tn{;} $\bs{a}_j\in\R^n$\tn{;} $\bs{b}_j\in\R^n_+$\tn{;} 
$c_j\in\R$\tn{;} $d_j\in\R$, $d_j>0$\tn{;} $s_j\in\R$, $s_j\geq 0$.\\[0.2cm]
Then the rule base is radial.
\end{theorem}

\noindent
\textbf{Proof.}
Having the $act$ function given by (\ref{actArch}), the specification
of the fuzzy sets according to (\ref{ArchDef}) coincides with the formulas 
(\ref{RadASet}) and (\ref{RadBSet}). What we need to show is that the radial
property (\ref{RadProp}) holds and the specification of the $act$ function
meets the condition (i) of Definition
\ref{raddef}.

Using the additive generator, we can write $T$ as $T(a,b)=t^{(-1)}(t(a)+t(b)).$
Let us show that we have also $T(a,T(b,c))=t^{(-1)}(t(a)+t(b)+t(c)).$

In order to do that, we recall from Section \ref{prelim} that
the pseudo-inverse of the continuous additive generator $t$ of an 
Archimedean $t$-norm is defined as
\begin{equation}
\label{psdefloc}
t^{(-1)}(z)=\left\{
\begin{array}{ll}
t^{-1}(z)&\;\;\;\tn{for}\;\;\;\;\;z\in[0,t(0)],\\
0&\;\;\;\tn{for}\;\;\;\;\;z\in(t(0),+\infty]\\
\end{array}
\right . 
\end{equation}
and therefore the following formulas hold:
\begin{equation}
\label{tt01}
\hspace{-2.4cm}
t^{(-1)}(t(z))=\,z\;\;\;\;\;\mr{for}\;\;\;\;\; z\in[0,1],
\end{equation}
\vspace{-0.3cm}
\begin{equation}
\label{tt02}
t(t^{(-1)}(z))=\left\{
\begin{array}{ll}
z&\;\;\;\tn{for}\;\;\;\;\;z\in[0,t(0)],\\
t(0)&\;\;\;\tn{for}\;\;\;\;\;z\in(t(0),+\infty].
\end{array}
\right .
\end{equation}

We have
$
T(a,T(b,c))=t^{(-1)}(\,t(a)+t(t^{(-1)}(t(b)+t(c)))\,)
$
and there are two cases possible:\\
1) $t(b)+t(c)\leq t(0)$. According to (\ref{tt02}),
$t(t^{(-1)}(t(b)+t(c)))=t(b)+t(c)$; and therefore
$T(a,T(b,c))=t^{(-1)}(t(a)+t(b)+t(c))$.\\
2) $t(b)+t(c)> t(0)$. By (\ref{tt02}),
$t(t^{(-1)}(t(b)+t(c)))=t(0)$; and therefore $T(a,T(b,c))=t^{(-1)}(t(a)+t(0))$.
Since $t(a)+t(0)\geq t(0)$, we obtain from (\ref{psdefloc})
$T(a,T(b,c))=0$. But
$t^{(-1)}(t(a)+t(b)+t(c))\!=0$ as $t(a)+t(b)+t(c)>t(0)$, hence
$T(a,T(b,c))=t^{(-1)}(t(a)+t(b)+t(c))$ as well.

By induction and the same approach 
it can be shown that for any $n\in\N$, $n>1$ and $u_i\in[0,1]$,
$i=1,\dots ,n$ it also holds
\begin{equation}
\label{TnormN}
T(u_1,\dots,u_n)=t^{(-1)}(t(u_1)+\dots +t(u_n)).
\end{equation}

Now, let the $act$ and $A_{ji}$ functions be specified according to (\ref{actArch})
and (\ref{ArchDef}). We denote $u_{ji}=(x_i-a_{ji})/b_{ji}$
for $i=1,\dots,n$, $\bs{a}_j=(a_{j1},\dots, a_{jn})$, $\bs{b}_j=(b_{j1},\dots,b_{jn})$.
Then, using (\ref{TnormN}), the representation of the antecedents $A_j(\bs{x})$
writes as
\begin{eqnarray}
A_j(\bs{x})&=&t^{(-1)}\left[\sum^n_{i=1} t(t^{(-1)}(q|u_{ji}|^p))\right].
\end{eqnarray}
By (\ref{tt02}), one reads the above as
\begin{equation}
\label{Aj3}
A_j(\bs{x})=t^{(-1)}\left[\sum^n_{i=1} \min\{t(0),q|u_{ji}|^p\}\right].
\end{equation}

We have again two cases possible: 1) If $q|u_{ji}|^p<t(0)$ for all~$i$,
then (\ref{Aj3}) has the form 
$A_j(\bs{x})=t^{(-1)}(\sum_{i=1}^n q|u_{ji}|^p)$, which
can be clearly written as
$A_j(\bs{x})=t^{(-1)}(q[\!\sqrt[p]{\sum_{i=1}^n |u_{ji}|^p}\,]^{^p})$, i.e,
$A_j(\bs{x})=act(||\bs{u}||_p)=act(||\bs{x}-\bs{a}_j||_{p,\bs{b}_j})$.

2) If there exists an $i$ such that $q|u_{ji}|^p\geq t(0)$, then, 
on the one hand, the sum in (\ref{Aj3}) is greater or equal to
$t(0)$ and therefore $A_j(\bs{x})=0$. On the other hand, 
$\sum_{i=1}^n q|u_{ji}|^p\geq t(0)$, i.e., 
$q[\!\sqrt[p]{\sum^n_i |u_{ji}|^p}\,]^{^p}=
q||\bs{x}-\bs{a}_j||^p_{p,\bs{b}_j}\geq t(0)$ and
therefore $act(||\bs{x}-\bs{a}_j||_{p,\bs{b}_j})=0$.
Thus, $A_j(\bs{x})=act(||\bs{x}-\bs{a}_j||_{p,{\bs{b}_j}})$
also in this case and we see that the specification of the $act$ function according
to (\ref{actArch}) is sufficient for the validity of the radial property.

In order to complete the proof, we must show that the function 
$t^{(-1)}(qz^p)$, $q>0$, $p\in[1,+\infty)$, $z\in[0,+\infty)$ can 
be considered as the $act$ function. For $q=1$, $p=1$ consider 
the properties of the continuous additive generator and its pseudo-inverse:

1) Either the value of $t(0)$ is bounded, i.e., $t(0)<+\infty$. Then
$t^{(-1)}$ is continuous, strictly decreasing on $[0,t(0)]$ and
equals to zero on $[t(0),+\infty]$. 
Thus, $t^{(-1)}$ is continuous and non-increasing on the 
interval $[0,+\infty]$.

2) Or the value of $t(0)$ is unbounded, i.e., $t(0)=+\infty$.
Then $t^{(-1)}$ is continuous, strictly decreasing on
$[0,+\infty]$ and $t^{(-1)}(t(0))=t^{(-1)}(+\infty)=0$,
which means that $\lim_{z\rightarrow+\infty} t^{(-1)}(z)=0$.

Hence $t^{(-1)}(z)$ meets the conditions of Defintion~\ref{raddef}
on the $act$ function. 

Since the function $qz^p$, $z\in[0,+\infty)$ is a continuous strictly increasing
bijection on $[0,+\infty)$ for $q>0$, $p\in[1,+\infty)$, the function 
$t^{(-1)}(qz^p)$ can be considered as the $act$ function too.\hfill$\Box$

The proved theorems give us tools for constructing the radial rule bases
and subsequently the \textit{radial fuzzy systems}. Several examples of 
these systems are presented below.

\subsection{Radial fuzzy systems}
\label{radSec}
Here we deliver the formal definitions of the radial fuzzy~systems. 
We will distinguish between the conjunctive and implicative radial fuzzy systems.
\begin{definition}
Let FS be the fuzzy system. We call it \textbf{the radial conjunctive fuzzy system}
(radial C-FS) if it is the conjunctive fuzzy system in the standard configuration and
its rule base is radial.
\end{definition}

\begin{definition}
Let FS be the fuzzy system. We call it \textbf{the radial implicative fuzzy system}
(radial I-FS) if it is the implicative fuzzy system in the standard configuration and
its rule base is radial.
\end{definition}

By inspecting the definitions, we see that the fuzzy system is radial if it is
in the standard configuration and has the radial rule base. The definitions
enable us to identify computational models of the radial fuzzy systems and
to study their properties. Before we discuss these computational models,
we present two concrete examples of the radial fuzzy systems. The first
system is based on the minimum $t$-norm and the other on the
product \mbox{$t$-norm}.

\subsubsection{Mamdani radial conjunctive fuzzy system}
\label{MamdrFS}
The name of this radial fuzzy system refers to the pioneering work of
Mamdani and Assilian who used the minimum $t$-norm to build a fuzzy
controller~\cite{Mamdani75}.

\begin{figure}[!htb]
\begin{center}
\includegraphics[width=5.3in]{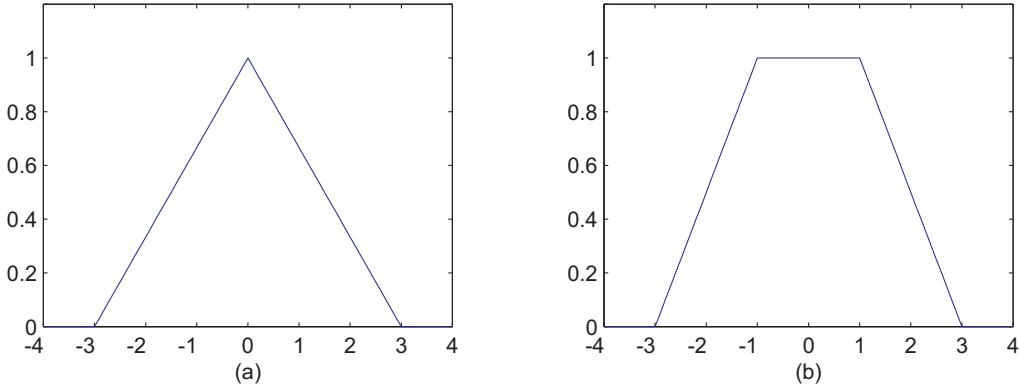}
\caption{The Mamdani radial C-FS. (a) An example of the antecedent fuzzy set 
for $a=0$, $b=3$; (b) An example of the consequent fuzzy set for $c=0$, $d=2$, $s=1$.}
\label{fig0401}
\end{center}
\end{figure}

The Mamdani radial C-FS use the minimum $t$-norm
$T_{\mr{M}}(a,b)=\min\{a,b\}$ and the triangular $act$ function:
$act(z)=\max\{0,1-z\}$, \mbox{$z\in[0,+\infty)$} which leads to the following 
membership functions of the respective fuzzy sets:
\begin{eqnarray*}
\label{MA}
A^M_{ji}(x_i)&=&\max \left\{0,1- \frac{|x_i-a_{ji}|}{b_{ji}} \right\},\\[0.2cm]
\label{MB}
B^M_j(y)&=&\max \left\{0,1-\frac{\max\{0,|y-c_j|-s_j\}}{d_j} \right\}
\end{eqnarray*}
where $\bs{a}_j\in\R^n$, $\bs{a}_j=(a_{j1},\dots,a_{jn})$;
$\bs{b}_j\in\R^n_+$, $\bs{b}_j=(b_{j1},\dots,b_{jn})$, $b_{ji}>0$;
$c_j\in\R$; $d_j\in\R$, $d_j>0$; $s_j\in\R$, $s_j\geq 0$ are parameters.
Examples of these membership functions
are presented in Fig.~\ref{fig0401}.

Clearly, the $act$ function $act(z)=\max\{0,1-z\}$, $z\in[0,+\infty)$
does satisfy the requirement (ia) of Definition \ref{raddef} for $z_0=1$.
Moreover, since the minimum $t$-norm is used,  the assumptions of 
Theorem~\ref{RadTheorem1} are fulfilled and the Mamdani C-FS is really
radial, i.e., it exhibits the radial property.

\begin{figure}[!htb]
\begin{center}
\includegraphics[width=5.3in]{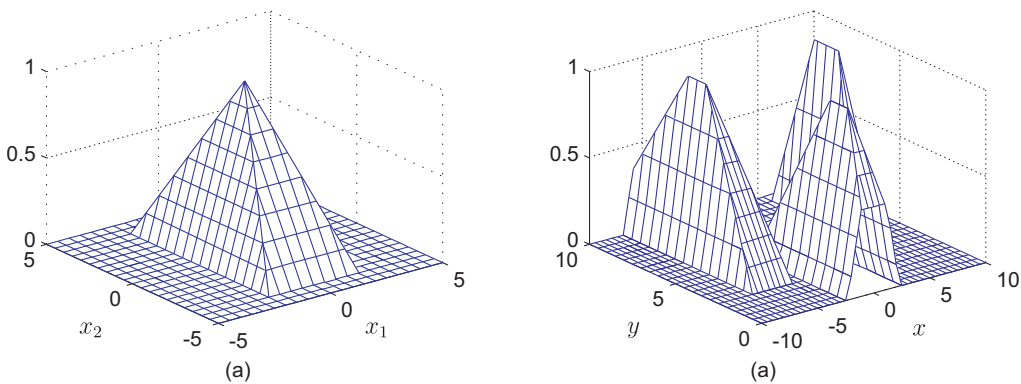}
\caption{The Mamdani radial C-FS. (a) An example of the antecedent for $n=2$,
$\bs{a}=(0,0)$, $\bs{b}=(2,4)$; (b) An example of the rule base for $n=2$, $m=3$.}
\label{fig0402ab}
\vspace{-0.5cm}
\end{center}
\end{figure}

Due to the radial property, the antecedent of the $j$-th rule in the Mamdani 
system writes
$$
A^M_j(\bs{x})=\min \{A^M_{j1}(x_1),\dots,A^M_{jn}(x_n)\}=
\max\{0,1-||\bs{x}-\bs{a}_j||_{\infty,\bs{b}_j}\}.
$$
Thus, the common $\ell_p$ norm is the cubic norm. Fig.~\ref{fig0402ab}
presents examples of the antecedent and the rule base in the Mamdani radial C-FS.

\subsubsection{Gaussian  radial implicative fuzzy system}
\label{GaussrFS}
This radial fuzzy system is considered as the implicative one. The system uses
the product $t$-norm $T_{\mr{P}}(a,b)=a\cdot b$ and the corresponding residuated
implication which is the Goguen implication. Recall
that this implication is specified as $a\ra_{\mr{P}}b= 1$ for $a\leq b$ and 
$a\ra_{\mr{P}}b=b/a$ for $a>b$, $a,b\in[0,1]$.

The product $t$-norm is continuous Archimedean.
The pseudo-inverse of its additive generator corresponds to the exponential function:
$t^{(-1)}(z)=\exp(-z)$ for $z\in[0,+\infty)$, $\exp(-\infty)=0$. Theorem 
\ref{RadTheorem2} is employed to obtain the specification of the fuzzy sets in 
the Gaussian radial I-FS. Using the above pseudo-inverse $t^{(-1)}$, setting
$q=1$ and choosing $p=2$, the membership functions of the respective fuzzy
sets writes as
\begin{eqnarray}
A^G_{ji}(x_i)&=&\exp\left[-\frac{(x_i-a_{ji})^2}{b_{ji}^2}\right],\\[0.1cm]
B^G_j(y)&=&\exp\left[-\frac{\max\{0,|\,y-c_j|-s_j\}^2}{d_j^2}\right].
\end{eqnarray}

\begin{figure}[!htb]
\begin{center}
\includegraphics[width=5.3in]{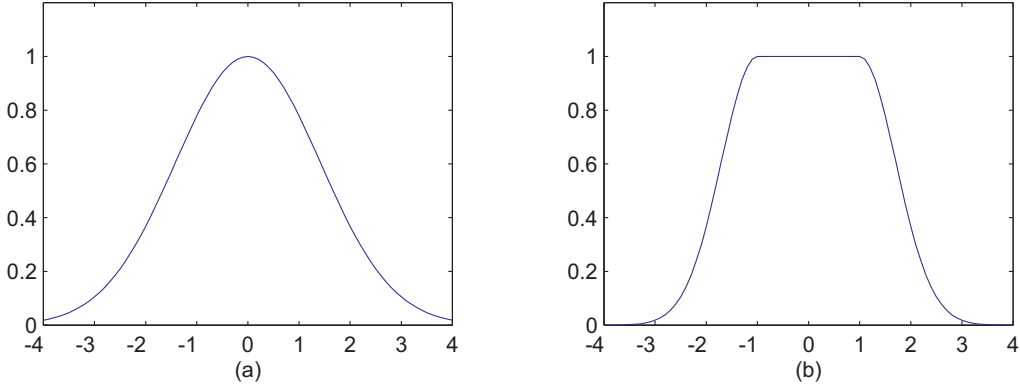}
\caption{The Gaussian radial I-FS. (a) An example of the antecedent fuzzy set
for $a=0$, $b=2$; (b) An example of the consequent fuzzy set for $c=0$, $d=1$, $s=1$.}
\label{fig0403}
\end{center}
\end{figure}

The $act$ function in the Gaussian radial I-FS system is determined by 
the exponential function in the form $act(z)=\exp(-z^2), z\in[0,\infty)$.
Hence the requirement (ib) of Definition \ref{raddef} applies 
and the assumptions of Theorem \ref{RadTheorem2} are fulfilled.

We can see that the membership functions of the antecedent fuzzy sets
$A^G_{ji}(x)$ coincide with the Gaussian curves. The membership functions
of the consequent fuzzy sets $B^G_j(y)$ are then given by their trapezoidal
modification. Fig.~\ref{fig0403} presents examples of both types of fuzzy sets.

By simple computation, the antecedent of the $j$-th rule in the Gaussian 
I-FS writes as
\begin{equation}
\label{antG}
A^G_j(\bs{x})=A^G_{j1}(x_1)\cdot\; \dots\; \cdot A^G_{jn}(x_n)=
\exp(-||\bs{x}-\bs{a}_j||_{2,{\bs{b}_j}}).
\end{equation}
Thus, the common $\ell_p$ norm is the Euclidean norm.

\begin{figure}[!htb]
\begin{center}
\includegraphics[width=5.3in]{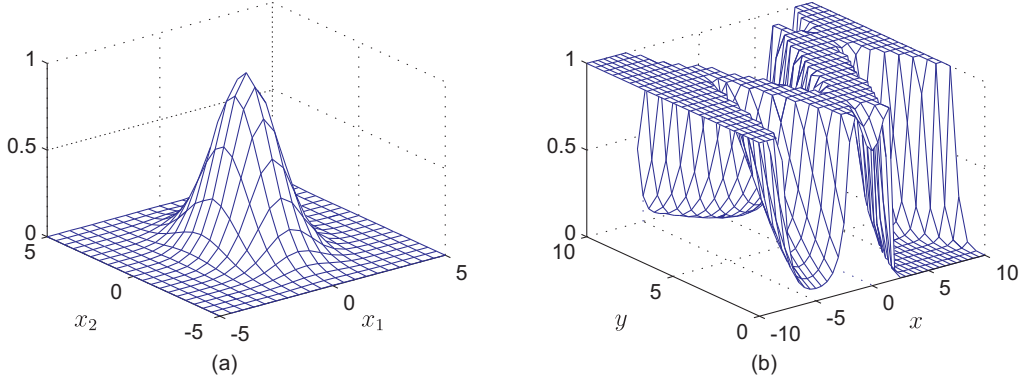}
\caption{The Gaussian radial I-FS. (a) An example of the antecedent for $n=2$, $\bs{a}=(0,0)$,
$\bs{b}=(2,4)$; (b) An example of the rule base for $n=1$, $m=3$.}
\label{fig0404ab}
\vspace{-0.5cm}
\end{center}
\end{figure}

In Fig.~\ref{fig0404ab}(a) an example of the antecedent is presented 
for $n=2$. In Fig.~\ref{fig0404ab}(b) there is presented the fuzzy relation
that corresponds to the rule base of the Gaussian radial I-FS that consists
of three rules.

\section{Computational models of radial fuzzy systems}
In this section we discuss the computational models of the radial conjunctive and 
implicative fuzzy systems. We show that the radial property makes the latter
mathematically tractable. We also investigate the mutual relationship between
both models. To proceed, let us stress that from now on we will use bare $\bs{x}$
to denote the input into the fuzzy system instead of the former $\bs{x}^*$.

\subsection{Computational model of radial conjunctive fuzzy systems}
The situation is rather straightforward here. Due to the radial property and
the radial character of the consequent fuzzy sets we have the following
modification of Lemma~\ref{LemmaC} for the radial conjunctive
fuzzy systems:
\begin{lemma}
Let $FS_{rconj}$ be the radial conjunctive fuzzy system. Let its rule base be complete. 
Then the fuzzy system computes the function
\begin{equation}
FS_{rconj}(\bs{x})=
\frac{\sum_{j=1}^m act(||\bs{x}-\bs{a}_j||_{p,\bs{b}_j})\cdot c_j}
{\sum_{j=1}^m act(||\bs{x}-\bs{a}_j||_{p,\bs{b}_j})}.
\label{FSrconj}
\end{equation}
\end{lemma}

\noindent
\textbf{Proof.}
To adapt the formula (\ref{FSconj}) of Lemma~\ref{LemmaC}, we put forward
two observations. First, the antecedents write
$A_j(\bs{x})=act(||\bs{x}-\bs{a}_j||_{p,\bs{b}_j})$ due to the radial property.
Second, the centroids of the consequents fuzzy sets $B_j$ correspond to
their central points $c_j$ because of the radial symmetry of the $B_j$ 
sets.\hfill$\Box$

\subsection{Computational model of radial implicative fuzzy systems}
In Section~\ref{FSstd}, we have shown that the computational model 
of the implicative fuzzy systems draws on the specification of the $\alpha$-cuts
\begin{equation}
\label{BAcut}
[B_j]^{A_j(\bs{x})}=\{y\,|\,A_j(\bs{x})\leq B_j(y)\}.
\end{equation}
Let us show that these $\alpha$-cuts can be stated explicitly in the case
of the radial implicative fuzzy systems.

\begin{definition}
For an input $\bs{x}\in X$, the output of the $j$-th rule 
of the radial implicative fuzzy system is the interval
\begin{equation}
\label{Ijdef}
I_j(\bs{x})=\left\{
\begin{array}{ll}
(-\infty,+\infty)&\;\;\;\mr{for}\;\;\;\;\;A_j(\bs{x})=0\\
I^+_j(\bs{x})&\;\;\;\mr{for}\;\;\;\;\;A_j(\bs{x})>0
\end{array}
\right.
\vspace{-0.3cm}
\end{equation}
where the interval
\begin{equation}
\label{Ij+def}
I^+_j(\bs{x})=[\;c_j-d_j||\bs{x}-\bs{a}_j||_{p,\bs{b}_j}-s_j\;,
\;c_j+d_j||\bs{x}-\bs{a}_j||_{p,\bs{b}_j}+s_j\;]
\end{equation}
is called the positive part of the output of the $j$-th rule for 
the input $\bs{x}\in X$.
\label{I_jdef}
\end{definition}

\begin{theorem}
\label{cutth}
In the radial I-FS,
for any $\bs{x}\in X$ and $j=1,\dots, m$, the $[B_j]^{A_j(\bs{x})}$
coincides with the output of the $j$-th rule, i.e.,
\begin{equation}
[B_j]^{A_j(\bs{x})}=I_j(\bs{x}).
\end{equation}
\end{theorem}

\noindent
\textbf{Proof.}
For a given $\bs{x}\in X$,
if $A_j(\bs{x})=0$, then $[B_j]^{0}=(-\infty,+\infty)$  by Definition \ref{alphaDef}
of the $\alpha$-cut.

If $A_j(\bs{x})>0$, then we have the respective $\alpha$-cut specified by the formula 
(\ref{BAcut}), which can be expanded into the following chain
of the equivalent inequalities:
\vspace{-0.1cm}
\begin{eqnarray}
A_j(\bs{x})&\leq&B_j(y),\label{st0}\nonumber\\
act(||\bs{x}-\bs{a}_j||_{p,\bs{b}_j})&\leq& act(\max\{0,|y-c_j|-s_j\}/d_j),\label{st1}\nonumber\\
||\bs{x}-\bs{a}_j||_{p,\bs{b}_j}&\geq& \max\{0,|y-c_j|-s_j\}/d_j,\label{st2}\nonumber\\
d_j||\bs{x}-\bs{a}_j||_{p,\bs{b}_j}&\geq& \max\{0,|y-c_j|-s_j\},\nonumber\\
d_j||\bs{x}-\bs{a}_j||_{p,\bs{b}_j}&\geq& |y-c_j|-s_j,\label{st4}\nonumber\\
d_j||\bs{x}-\bs{a}_j||_{p,\bs{b}_j}+s_j&\geq& |y-c_j|\label{st5}.
\end{eqnarray}

In the chain, only the step from the second to the third inequality should
be explained. By Definition~\ref{raddef}, the $act$ function is generally non-increasing
and strictly decreasing on interval $[0,z_0)$ for some
$z_0\in(0,\infty]$. Further, we have $act(z_+)>0$ for $z_+\in[0,z_0)$. Hence,
if $0<act(z_+)\leq act(z)$, then it must be that $z_+\geq z$. 
As we have assumed that $0<act(||\bs{x}-\bs{a}_j||_{p,\bs{b}_j})$, 
the discussed step is correct.~\hfill$\Box$


The inequality (\ref{st5}) determines the closed interval which can be considered
as the output of the $j$-th rule for the input $\bs{x}\in X$
if $A_j(\bs{x})>0$. It is this fact that is reflected in Definition \ref{I_jdef}
when labeling the interval $I^+_j(\bs{x})$. Note that the limit points
of $I^+_j(\bs{x})$ depend on all parameters 
$\bs{a}_j, \bs{b}_j, c_j, d_j$ and $s_j$.

To proceed, we prove a simple lemma dealing with the intersection of closed
intervals in $\R$.

\begin{lemma}
Let $\{I_j\}_{j=1}^m$ be a set of $\,m$ non-empty closed intervals in $\R$.
Let $L(I_j)$ or $R(I_j)$ denotes the left or the right limit point of the $j$-th 
interval, respectively, i.e., $I_j=[L(I_j),R(I_j)]$.
Then the intersection $\bigcap_{j=1}^m I_j$ is non-empty if and only if
\begin{equation}
\label{condefcond}
\max_j \{L(I_j) \}\le \min_j \{R(I_j) \};
\end{equation}
and we have
\begin{equation}
\bigcap_{j=1}^m I_j=[\max_j \{L(I_j) \}\,,\,\min_j \{R(I_j)\}].
\end{equation}
\label{intLemma}
\end{lemma}

\vspace*{-1ex}
\noindent
\textbf{Proof.}
Because we work with the closed intervals in $\R$, each interval writes as
$I_j=[L(I_j),+\infty)\cap (-\infty,R(I_j)]$.  The intersection of $m$ 
intervals is then given~as
$$
\bigcap_{j=1}^m I_j=
[L(I_1),+\infty)\cap (-\infty,R(I_1)]
\cap\;\dots\;\cap
[L(I_m),+\infty)\cap (-\infty,R(I_m)].
$$
Due to the commutativy and associativity of the intersection,
the right-hand side has the form
$$
(\;[L(I_1),+\infty)\,\cap \dots \cap\, [L(I_m),+\infty)\;)
\cap \dots \cap
(\; (-\infty,R(I_1)]\,\cap \dots \cap\, (-\infty,R(I_m)]\;)
$$
and therefore
$
\bigcap_{j=1}^m I_j=
[\max_j \{ L(I_j) \},+\infty ) \cap (-\infty, \min_j \{ R(I_j) \}].
$
This gives a non-empty set if and only if 
$
\max_j \{ L(I_j) \}\leq  \min_j \{ R(I_j) \};
$
and in this case
$$
\bigcap_{j=1}^m I_j=
[\max_j \{ L(I_j) \}\,,\,\min_j \{ R(I_j) \}].
$$
This finishes the proof.\hfill$\Box$

\begin{lemma}
Let $FS_{rimpl}$ be the coherent radial implicative fuzzy system.
Let its rule base be complete. Then the fuzzy system computes the function
\begin{equation}
FS_{rimpl}(\bs{x})=\frac{\max_{j|A_j(\bs{x})>0}\{L(I^+_j(\bs{x})\}+
\min_{j|A_j(\bs{x})>0}\{R(I^+_j(\bs{x}))\}}{2}.
\end{equation}
\end{lemma}

\noindent
\textbf{Proof.} As we assume that the rule base of the fuzzy system is complete,
the set of indices $\{j|A_j(\bs{x})>0\}$ is non-empty for any input $\bs{x}\in X$.
Further, because of completeness and coherence we have also
$$
\bigcap_{j|A_j(\bs{x})>0} I^+_j(\bs{x})=\bigcap_{j=1}^m I_j(\bs{x})
\not=\emptyset
$$ for any input $\bs{x}\in X$. Definition~\ref{I_jdef} and
Lemma~\ref{intLemma} give us the specification of the left-hand side
of the above equality, i.e., 
$$
\bigcap_{j|A_j(\bs{x})>0} I^+_j(\bs{x})=
[\max_{j|A_j(\bs{x})>0}\{L(I^+_j(\bs{x})\},
\min_{j|A_j(\bs{x})>0}\{R(I^+_j(\bs{x}))\}\,].
$$
According to Theorem~\ref{cutth}, the right-hand side corresponds to the 
intersection of the $[B_j]^{A_j(\bs{x})}$ $\alpha$-cuts, which are the cores
of the fuzzy sets issued from the inference engine, and the assertion is obtained
by Lemma~\ref{lemma2}.\hfill$\Box$

\subsection{Relation between computational models}
The following lemma tells us that under the assumptions of completeness
and coherence, the computations of the above models cannot differ
arbitrarily much.

\begin{lemma}
Let RB be the complete radial rule base. Let $FS_{rconj}$ and $FS_{rimpl}$ be 
the radial fuzzy systems based on this RB. Let $FS_{rimpl}$ be coherent, then 
$$
|FS_{rconj}(\bs{x})-FS_{rimpl}(\bs{x})|<(c_{max}-c_{min})
$$
for every $\bs{x}\in X$ where $c_{max}=\max_{j=1}^m\{c_j\}$ and 
$c_{min}=\min_{j=1}^m\{c_j\}$.
\end{lemma}

\noindent
\textbf{Proof.} $c_j$ are the centroids of the consequent fuzzy sets $B_j$, 
$j=1,\dots,m$. First, we have $FS_{rconj}(\bs{x})\in [c_{min},c_{max}]$,
because the output of $FS_{rconj}$ is given as a weighted average of $c_j$;
and it is well known that the weighted average lies in-between the minima
and maxima of the weighted points.

We further argue that also $FS_{rimpl}(\bs{x})\in [c_{min},c_{max}]$ for
any $\bs{x}\in X$. Let $J^+=\{j|A_j(\bs{x})>0\}$ and
$I=\bigcap_{j\in J^+} I_j^+(\bs{x})=
[y_{min},y_{max}]\not=\emptyset$.
Either $I\subseteq [c_{min},c_{max}]$ and we are done or there exists
a point from $I$ that lies outside the interval $[c_{min},c_{max}]$. Let
$y_{max}>c_{max}$, we have $y_{max}\in I^+_j$ for all~$j\in J^+$
because $I$ is given by the intersection of the $I^+_j$ intervals. Now,
let $r_{max}=y_{max}-c_{max}>0$. We have also 
$c_{max}-r_{max}\in I$. Indeed, let $r_{j,max}=y_{max}-c_j$, 
then $c_j+r_{j,max}=y_{max}\in I^+_j$, $j\in J^+$.
As $I^+_j$ intervals are symmetric around the central
points $c_j$, one has also $c_j-r_{j,max}\in I^+_j$, $j\in J^+$.
Further $r_{j,max}\geq r_{max}$ for each $j\in J^+$, so also
$c_j-r_{max}\in I^+_j$ for each $j\in J^+$. Because 
$c_j\leq c_{max}<y_{max}$, and $y_{max}\in I^+_j$, we have
also $c_j-r_{max}\leq c_{max}-r_{max}<y_{max}-r_{max}=
c_{max}<y_{max}$ for all $j\in J^+$; and therefore 
$c_{max}-r_{max}\in I$.

Because $c_{max}-r_{max}\in I$, we have $y_{min}\leq c_{max}-r_{\max}$,
so $y_{min}+y_{max}\leq c_{max}-r_{max}+r_{max}+c_{max}$,
i.e., $y^*=(y_{min}+y_{max})/2\leq c_{max}$. Now, if $c_{min}\leq y_{min}$
then we are done. If $y_{min}<c_{min}$, then we get in the same way as
above that $c_{min}+r_{min}\in I$ for $r_{min}=c_{min}-y_{min}$. This
gives $y_{max}\geq c_{min}+r_{min}$, $y^*=(y_{min}+y_{max})/2\geq c_{min}$
and we finally have $y^*\in [c_{min},c_{max}]$.\hfill$\Box$

\section{Coherence of radial implicative fuzzy systems}
\label{cohSec}
In this section, we deal with the coherence of the radial implicative fuzzy systems.
The main result presented here is the specification of a sufficient condition
for a radial I-FS to be coherent. This sufficient condition depends
only on the parameters of the system. We start by two straightforward 
lemmas.

\begin{lemma}
\label{rct0.5}
The radial I-FS is coherent if and only if for any input $\bs{x}\in X$
the intersection of the outputs of its rules is non-empty, i.e., if for every 
$\bs{x}\in X$
\begin{equation}
\label{ii0.5}
\bigcap_{j=1}^m I_j(\bs{x})\not=\emptyset.
\end{equation}
\end{lemma}

\noindent
\textbf{Proof.}
The assertion is a direct corollary of Theorems \ref{ctbasic} and 
\ref{cutth}.~\hfill$\Box$

Since we always have $I^+_j(\bs{x})\subseteq I_j(\bs{x})$, no matter
what the value of $A_j(\bs{x})$~is, we can formulate the following
more convenient sufficient condition for the radial I-FS to be coherent.

\begin{lemma}
\label{rct01}
In the radial I-FS, if the intersection of the positive parts of the outputs of rules 
is non-empty for any input $\bs{x}\in X$, i.e., if
\begin{equation}
\label{ii}
\bigcap_{j=1}^m I^+_j(\bs{x})\not=\emptyset
\end{equation}
for every $\bs{x}\in X$, then the fuzzy system is coherent.
\end{lemma}

\noindent
\textbf{Proof.}
If $A_j(\bs{x})=0$ for a given $\bs{x}$ and $j$,  then
$I^+_j(\bs{x})\subset I_j(\bs{x})=(-\infty,+\infty)$.
If $A_j(\bs{x})>0$, then $I^+_j(\bs{x})=I_j(\bs{x})$.
Hence for any $\bs{x}\in X$ and $j\in\{1,\dots,m\}$
one has
$
I^+_j(\bs{x})\subseteq I_j(\bs{x})
$
and therefore
$$
\bigcap_{j=1}^m I^+_j(\bs{x})
\subseteq 
\bigcap_{j=1}^m I_j(\bs{x}).
$$
If $\bigcap_{j=1}^m I^+_j(\bs{x})\not=\emptyset$,
then also 
$\bigcap_{j=1}^m I_j(\bs{x})\not=\emptyset$ for $\bs{x}\in X$
and we get the result by Lemma~\ref{rct0.5}.
\hfill$\Box$

By Lemma~\ref{rct01}, we have transformed the coherence question
into the testing the non-emptiness of the intersection of closed intervals,
which is more specific than testing the intersection of more general
$\alpha$-cuts. But the theorem is still weak in the sense that the 
intersection has to be checked for any input  $\bs{x}\in X\subseteq\R^n$.
In the rest of the section, we transform the problem further into 
testing only relationships among the parameters of the fuzzy system.

To simplify the notation a bit, we will write $||\cdot||_{\bs{b}_j}$
instead of $||\cdot||_{p,\bs{b}_j}$ and $||\cdot||$ instead of $||\cdot||_p$,
i.e., we omit the $p$ index keeping in mind that we are working with 
the $\ell_p$ norms.

\begin{theorem}
\label{cth}
Let the radial implicative fuzzy system consists of $m$ rules.
If for any pair of rules $j,k\in\{1,\dots,m\}$
\begin{equation}
\label{ncc}
|c_j-c_k|-(s_j+s_k)\;\leq\; \min\{d_j\alpha_j,d_k\alpha_k\}\cdot ||\bs{a}_j-\bs{a}_k||
\vspace{0.2cm}
\end{equation}
where $\alpha_j=1/\max_i\{b_{ji}\}$, $\alpha_k=1/\max_i\{b_{ki}\}$,
then the radial I-FS is coherent.
\end{theorem}

\noindent
\textbf{Proof.}
By Definition \ref{I_jdef}, the positive part of the output of the $j$-th rule
is specified for an input $\bs{x}\in X$ as
\begin{equation}
\label{infint}
I^+_j(\bs{x})=[\;c_j-d_j||\bs{x}-\bs{a}_j||_{\bs{b}_j}-s_j\;,\;
c_j+d_j||\bs{x}-\bs{a}_j||_{\bs{b}_j}+s_j\;].
\end{equation}
Having $m$ rules in the rule base, we are given $m$ intervals $I^+_j(\bs{x})$,
$\bs{x}\in X$.
We write them as $I^+_j(\bs{x})=[L(I^+_j(\bs{x})),R(I^+_j(\bs{x}))]$ where
$$
L(I^+_j(\bs{x}))=c_j-d_j||\bs{x}-\bs{a}_j||_{\bs{b}_j}-s_j\,\;\mathrm{and}\,\;
R(I^+_j(\bs{x}))=c_j+d_j||\bs{x}-\bs{a}_j||_{\bs{b}_j}+s_j.
$$

According to Lemma~\ref{rct01}, a sufficient condition for the radial \mb{I-FS}
to be coherent is that $\bigcap_{j=1}^m I^+_j(\bs{x})\not=\emptyset$
for every $\bs{x}\in X$. Hence to prove our theorem it is sufficient to show that
``if (\ref{ncc}) holds for all pairs of rules $j,k\in\{1,\dots,m\}$, then (\ref{ii})
holds for every $\bs{x}\in X$ too". This is equivalent to ``if (\ref{ii}) does not hold
for some $\bs{x}^*\in X$, then also (\ref{ncc}) does not hold for some pair $j,k$''.
This is what we are going to prove.

Let (\ref{ii}) do not hold. Then, by Lemma \ref{intLemma}, there exists $j,k\in \{1,\dots ,m \}$,
$j\not=k$ such that $L(I^+_j(\bs{x}^*))>R(I^+_k(\bs{x}^*))$ for some $\bs{x}^*\in X$, i.e., 
$$
c_j-d_j||\bs{x}^*-\bs{a}_j||_{\bs{b}_j}-s_j>
c_k+d_k||\bs{x}^*-\bs{a}_k||_{\bs{b}_k}+s_k,
$$
which is
$$
c_j-c_k>d_j||\bs{x}^*-\bs{a}_j||_{\bs{b}_j}+
d_k||\bs{x}^*-\bs{a}_k||_{\bs{b}_k}+s_j+s_k.
$$

The last inequality is possible only if $(c_j-c_k)>0$, because the right-hand side
is always non-negative. Hence
\begin{equation}
\label{Jner}
|c_j-c_k|>d_j||\bs{x}^*-\bs{a}_j||_{\bs{b}_j}+d_k||\bs{x}^*-\bs{a}_k||_{\bs{b}_k}+s_j+s_k.
\end{equation}

It is straithforward to observe that for the given scaled $\ell_p$ norm $||\cdot||_{\bs{b}}$
and its unscaled version $||\cdot||$ there exists a positive number $\alpha$ such that
$\alpha||\bs{u}||\leq ||\cdot||_{\bs{b}}$ for any $\bs{u}\in\R^n$. 
Namely, $\alpha=1/\max\{b_i\}$. Setting $\alpha_j=1/\max_i\{b_{ji}\}$,
$\alpha_k=1/\max_i\{b_{ki}\}$, we have
$
\alpha_j||\bs{u}||\leq ||\bs{u}||_{\bs{b}_j},
\alpha_k||\bs{u}||\leq ||\bs{u}||_{\bs{b}_k}
$
for every $\bs{u}\in\R^n$.
Denoting
\begin{eqnarray}
\label{Jbjk}
J_{\bs{b}_{jk}}(\bs{x})&=&d_j||\bs{x}-\bs{a}_j||_{\bs{b}_j}+d_k||\bs{x}-\bs{a}_k||_{\bs{b}_k},\\
\label{Jjk}
J_{jk}(\bs{x})&=&d_j\alpha_j||\bs{x}-\bs{a}_j||+d_k\alpha_k||\bs{x}-\bs{a}_k||\nonumber
\end{eqnarray}
we get
$
J_{\bs{b}_{jk}}(\bs{x})\geq J_{jk}(\bs{x})
$
for any $\bs{x}\in\R^n$.

Let us search for the minimum of the $J_{jk}(\bs{x})$ with respect to $\bs{x}\in\R^n$.
By the triangle inequality for $||\cdot ||$, one has
$
||\bs{x}-\bs{a}_j||+||\bs{x}-\bs{a}_k||\geq ||\bs{a}_j-\bs{a}_k||
$
and therefore the minimum of the left-hand side is reached
either for $\bs{x}=\bs{a}_j$ or $\bs{x}=\bs{a}_k$, i.e.,
$
\min_{\bs{x}\,\in\,\R^n} \{||\bs{x}-\bs{a}_j||+||\bs{x}-\bs{a}_k||\}=||\bs{a}_j-\bs{a}_k||.
$\\

Concerning the minimum of $J_{jk}(\bs{x})$, there are two cases possible:

\bigskip
\noindent
1) $d_j\alpha_j\leq d_k\alpha_k$. Then
\begin{eqnarray*}
J_{jk}(\bs{x})&=&d_j\alpha_j\cdot (||\bs{x}-\bs{a}_j|| + ||\bs{x}-\bs{a}_k||)
+(d_k\alpha_k-d_j\alpha_j)\cdot ||\bs{x}-\bs{a}_k||,\\
\min_{\bs{x}\,\in\,\R^n}\{J_{jk}(\bs{x})\}&=&d_j\alpha_j||\bs{a}_k-\bs{a}_j||.
\end{eqnarray*}

\noindent
2) $d_j\alpha_j> d_k\alpha_k$. Then
\begin{eqnarray*}
J_{jk}(\bs{x})&=&d_k\alpha_k\cdot (||\bs{x}-\bs{a}_j|| + ||\bs{x}-\bs{a}_k||)
+(d_j\alpha_j-d_k\alpha_k)\cdot ||\bs{x}-\bs{a}_j||,\\
\min_{\bs{x}\,\in\,\R^n}\{J_{jk}(\bs{x})\}&=&d_k\alpha_k||\bs{a}_j-\bs{a}_k||.
\end{eqnarray*}

Combining both cases we obtain
\begin{equation}
\label{Jklmin}
\min_{\bs{x}\in\R^n}\{J_{jk}(\bs{x})\}=\min\{d_j\alpha_j,d_k\alpha_k\}\cdot 
||\bs{a}_j-\bs{a}_k||.
\end{equation}

The inequality (\ref{Jner}) holds for some $\bs{x}^*\in X$, hence it must
hold also for~$\bs{x}$ at which the right-hand side reaches its minimum
over $\R^n$. Thus using (\ref{Jbjk}), it follows from (\ref{Jner}) that
\begin{equation}
\label{cjkJbjk}
|c_j-c_k|>\min_{\bs{x}\,\in\,\R^n} \{J_{\bs{b}_{jk}}(\bs{x})\} +s_j+s_k.
\end{equation}
Since $J_{\bs{b}_{jk}}(\bs{x})\geq J_{jk}(\bs{x})$, we have
$
|c_j-c_k|>\min_{\bs{x}\,\in\,\R^n} \{J_{jk}(\bs{x})\} +s_j+s_k
$
as well, which by (\ref{Jklmin}) gives
$$
|c_j-c_k|>\min\{d_j\alpha_j,d_k\alpha_k\}\cdot ||\bs{a}_j-\bs{a}_k||+s_j+s_k\,.
$$
Rearranging the above we get
\begin{equation}
\label{wkl_impl}
|c_j-c_k|-(s_j+s_k)>\min\{d_j\alpha_j,d_k\alpha_k\}\cdot ||\bs{a}_j-\bs{a}_k||
\end{equation}
which was to prove.\hfill$\Box$\\

Let us comment on the theorem. What we have proved is that we are able 
to assure the coherence of the radial I-FS only by testing relations among
values of its parameters. This substantially improves Lemma~\ref{rct01}
because we have excluded the dependency of the coherence condition
on the values of the inputs to the radial I-FS.

By analyzing the formula (\ref{wkl_impl}), we see that if the $\min\{d_j\alpha_j,d_k\alpha_k\}$
is fixed, then the inequality is validated by increasing $|c_j-c_k|$ or decreasing 
\mbox{$||\bs{a}_j-\bs{a}_k||$}. We interpret this fact that if the radial I-FS is not coherent, then
there exists a pair of rules such that the centers of the antecedents are close, but the centers
of their consequents are distant. This corresponds to the intuitive meaning of the incoherence
that rules having similar preconditions state contrary conclusions \citep{Driankov93}.

In order to state the coherence according to Theorem~\ref{cth},
$(m^2-m)/2$ inequalities (\ref{ncc}) have to be verified. Indeed,
the full number of the inequalities is $m^2$, but they are symmetric
with respect to $j,k$ and the inequality trivially holds for $j=k$.


In order to finish the section, we will compare our results with those
of Dubois, Prade and Ughetto presented in \citep{Dubois97}. 
Section~IV of their paper, which concerns coherence of a set of 
parallel gradual rules \cite{Dubois96}, is relevant to our work.

In \citep{Dubois97}, the authors present three necessary and sufficient conditions
for an I-FS to be coherent: Propositions~4.1, 4.4 and 4.10. Proposition
4.1 methodologically corresponds to our Theorem~\ref{ctbasic}, and is 
practically inapplicable because it requires checking infinitely many
conditions in order to state the coherence of the I-FS.

Proposition~4.4 says that, if the I-FS consists of two rules and its
consequent fuzzy sets are convex, then one can form for each rule 
two functions of the input~$\bs{x}$ that delimit lower and upper
bounds of the cores of the relations that correspond to the individual rules. 
Denoting these functions $L_1,R_1$ or $L_2,R_2$ for the first or
the second rule, respectively, the I-FS is coherent if and only
if $R_1(\bs{x})\geq L_2(\bs{x})$ and $R_2(\bs{x})\geq L_1(\bs{x})$
for every input $\bs{x}\in \mr{supp}(A_1)\,\cap\,\mr{supp}(A_2)$.
This statement is rather general because an explicit specification of the
bounds may be hard and their comparison must be done for every 
input $\bs{x}\in \mr{supp}(A_1)\,\cap\,\mr{supp}(A_2)$.
Special cases simplify the situation and several algorithms performing
this checking are presented in the paper, but only for the pairs of SISO
(single-input single-output) rules that employ the trapezoidal fuzzy sets.

Concerning coherence of the MISO rules, in~Proposition~4.8 
Dubois~et~al. present a sufficient condition for coherence 
of two MISO rules. The proposition says that,
if there exists at least one $i\in\{1,\dots,n\}$, such that two
SISO rules $A_{1i}\rightarrow B_1$, $A_{2i}\rightarrow B_2$
are coherent, then two MISO rules are coherent as well.
$A_{1i}$, $A_{2i}$ are considered to be the components of 
the respective antecedents $A_1$, $A_2$ according to formula
(\ref{RuleAnt}). The authors point out that this is only 
a sufficient condition which is not necessary.

Comparing this result with our, we can say that our sufficient
condition is more specific than the one presented in Proposition~4.8.
As an example, consider two rules of the Gaussian radial I-FS, see
Section~\ref{GaussrFS}, with the parameters presented in Table~\ref{Tab4.1}.
The rules are pairwise coherent according to Theorem~\ref{cth}, 
but they are not coherent according to Proposition~4.8 in \citep{Dubois97}.

\begin{table}[htb]
\begin{center}
\begin{tabular}{c||c|c|c|c|c|c|c}
&$a_{j1}$&$a_{j2}$&$b_{j1}$&$b_{j2}$&$c_j$&$d_j$&$s_j$\\\hline\hline
$j=1$&0&0&1&1&0&1&0\\\hline
$j=2$&3&4&1&1&5&1&0 
\end{tabular}
\end{center}
\caption{Parameters of the Gaussian radial I-FS.}
\label{Tab4.1}
\end{table}

Proposition 4.10 states that the I-FS is coherent if and only if
its rules are pairwise coherent. This is reflected by our Theorem~\ref{cth}
because checking the inequality (\ref{ncc}) corresponds to the coherence test
for the pairs of rules $j,k\in\{1,\dots,m\}$. That is, if (\ref{ncc}) holds, then 
we are certain that the intersection $I^+_j(\bs{x})\,\cap\, I^+_k(\bs{x})$
is non-empty for any input $\bs{x}\in X$. Since intervals $I_j^+(\bs{x})$
are subsets of intervals $I_j(\bs{x})$, see Definition~\ref{I_jdef}, we are
assured by the validity of (\ref{ncc}) that $I_j(\bs{x})\cap I_k(\bs{x})\not=\emptyset$ 
for any input $\bs{x}\in X$. If this holds for all possible pairs of rules, we can imply
the validity of (\ref{ii0.5}), which gives the coherence of the radial I-FS.
This actually corresponds to the sufficiency part of Proposition~4.10.

Discussing the necessary condition for coherence, we point out the 
following two observations. In fact, there are two sources of why (\ref{ncc})
is not the necessary condition as well. First is the assertion of Lemma~\ref{rct01} 
that cannot be generally reversed. However, there is the special case
when this can be done. It is the case when the radial I-FS
employs the $act$ function of the (ib) type  in Definition~\ref{raddef}.  
The Gausian radial I-FS is an example of such the system. Then 
$I^+_j(\bs{x})=I_j(\bs{x})$ for all $j$ and $\bs{x}\in X$; and the 
non-emptiness of the intersection (\ref{ii}) is also necessary for 
coherence of the system.

The second source of sufficiency can be identified in the proof of 
Theorem~\ref{cth}. Namely, it is the replacement
of $J_{\bs{b}_{jk}}(\bs{x})$ by $J_{jk}(\bs{x})$.
If we did not take this replacement, then the inequality (\ref{cjkJbjk})
would read
$$
|c_j-c_k|-(s_j+s_k)>\min_{\bs{x}\,\in\,X} \{J_{\bs{b}_{jk}}(\bs{x})\}.
$$
The problem here is with an analytic specification of the minimum of 
$J_{\bs{b}_{jk}}(\bs{x})$ that must be searched numerically.
On the other hand, $J_{\bs{b}_{jk}}(\bs{x})$, $\bs{x}\in\,\R^n$
forms a~convex function and therefore numerical algorithms
are effective here.

To end the discussion, we state that the paper \citep{Dubois97} presents
algorithms for checking the coherence of two rules with the trapezoidal
fuzzy sets, but these are rather complicated and specific. Our 
Theorem~\ref{cth} is general with respect to the entire class of the radial
implicative fuzzy systems. In fact, we have enhanced the results and
methodology presented in \citep{Dubois97} mainly in practical applicability.

\section{Conclusions}
In the paper, we have introduced the class of the radial fuzzy systems. The class 
consists of the radial conjunctive and the radial implicative fuzzy systems. 
Computational models of the radial fuzzy systems draw on the notion of the radial
MISO rule base and the standard configuration of the fuzzy system. The standard
configuration comprises the singleton fuzzifier, the MISO rule base, the CRI inference
engine and the WA or MOM method of defuzzification. The WA method
is used in the conjunctive systems and the MOM method in the implicative
ones.

In the radial rule bases, fuzzy sets are implemented by the radial functions;
and  the antecedents of rules exhibit the radial property which is a~kind of 
the shape preservation property. In the paper, we have proved two theorems
that enable building the radial fuzzy systems on the basis of the minimum
$t$-norm and the continuous Archimedean $t$-norms.

Radial fuzzy systems compute functions according to their computational
models. We have specified these models explicitly and showed that the
radial property enables their convenient representations. With respect to the
computational models, the notions of completeness and coherence
were recalled. Completeness resides on the notion of the degree of 
covering (DOC) of the rule base. The DOC corresponds to the lower bound on
maximum of degrees of firing of individual rules when going across of the
input set of the fuzzy system.

Completeness ($\mathrm{DOC}\!>\!0$) is important not only to have the 
computational models of the radial fuzzy systems specified, but it may also
be shown that the DOC makes a lower bound when comparing the inference
engine's output fuzzy sets from the radial conjunctive and implicative fuzzy
systems~\cite{Coufal2015-EANN15}. We did not present here an analysis of
how the DOC can be computed for the radial rule bases. We only note that this
task falls into the area of computational geometry and is also related to the
problem of redundant rules detection~\cite{Dubois97, Stepnicka2015}.

The other important notion is coherence. First of all, remark that
coherence is not affected by the used defuzzification method. It relates to the
core of the inference engine's output fuzzy set. We have shown that for the radial
implicative fuzzy systems coherence can be assured by checking
a rather simple sufficient condition. The condition has the form of 
inequalities between the parameters of the IF-THEN rules that constitute
the rule base of the system. There is also the natural question on the necessary
condition on coherence. In fact, this is related to the global minimum of the
$J_{\bs{b}_{jk}}(\bs{x})$ function that was introduced in the proof 
of Theorem~\ref{cth}. This minimum must be searched numerically, but the
$J_{\bs{b}_{jk}}$ function is convex so local minima are also global minima
and numerical optimization works effectively here.

We see several application domains of the radial fuzzy systems. First, the input and 
output sets may be any normed spaces, hence elements of $X$ may be rather
complex objects, e.g., text documents. If we are able to introduce norms on these
spaces, the radial fuzzy systems can be used to represent functions from $X$ to~$Y$.
The implicative radial fuzzy systems then deliver a~logical structure into the functions
they represent.

Second, the radial fuzzy systems enable us to establish a bridge between the data
driven descriptions of relations that correspond to the conjunctive systems
and the logically driven descriptions that correspond to the implicative systems.
In fact, if we omit denominator in the WA method of defuzzification, then we
will find that computation of the radial C-FS corresponds to the computation
of a RBF neural network \cite{Haykin08}. The Gaussian systems are of priority
interest here \cite{ANFIS93, Kurkova2014}. Hence, there is an interesting field
of future research that deals with fusion of data driven learning algorithms
such as back-propagation and logical descriptions of relations. Moreover, using 
the kernel methods popular in machine learning \cite{Kung2014} might be fruitful
in the context of the radial fuzzy systems.

To sum up the novelties presented in the paper, we mention especially the actual
introduction of the class of the radial fuzzy systems. The class accommodates the 
fuzzy systems which are commonly used in applications~\cite{Handbook-ORANGE}.
Recall the Mamdani conjunctive fuzzy system of Section~\ref{MamdrFS} that
employs the triangular fuzzy sets and the minimum $t$-norm. The Gaussian
fuzzy sets and the product $t$-norm are also broadly used in applications
and the Gaussian conjunctive radial fuzzy system represents a popular choice
in practice \cite{Wang92, ANFIS93}. Hence the class as a whole does not only
represent a convenient theoretical concept, but it has also a~strong link to
real-world problems. Recall for instance the car navigation example of
Section~\ref{motiv}.

Furthermore, a broad variety of other members of the class can  be introduced
drawing on different Archimedean $t$-norms, see Appendix in \cite{Mesiar00}
for a~comprehensive list of various parameterized families of $t$-norms. Each
continuous Archimedean $t$-norm generates a radial rule base and consequently
a conjunctive or implicative radial fuzzy system. Hence an user can design the radial
fuzzy system that fits best her or his requirements. All such systems exhibit
the radial property and therefore admit a unified approach for analyzing their 
properties such as completeness and/or coherence of their rule bases.

Coherence is the most important from the properties of the radial implicative fuzzy
systems that we have analyzed in the paper. It corresponds to non-contradictoriness
of knowledge stored in an implicative fuzzy system. It~has been studied in different
contexts in literature, e.g.~in \cite{Yager91, Leung93, Viaene00}
and most thoroughly in \cite{Dubois97}, however, there has not been yet introduced 
a comfortable unifying view for its analysis. The framework of the radial fuzzy systems
enables establishing such a view due to Theorem 6 that state the sufficient condition
for testing coherence of any radial implicative fuzzy system. This makes a great benefit
because the user is liberated from analyzing properties of individual fuzzy systems
determined by individual shapes of the employed fuzzy sets.

Finally, we mention that the entire framework of the radial fuzzy systems seems
promising to bring more interesting results in several other directions. These are
related to the questions on universal approximation capabilities of the radial
fuzzy systems, detection of redundant rules in radial rule bases and connection
to the radial basis neural networks to mention a few. We are going to focus
on these directions in our future research.

\paragraph*{Acknowledgements} This work was supported by the COST grant 
LD13002 provided by the Ministry of Education, Youth and Sports of 
the Czech Republic.\\

\noindent
\textbf{References}




\bibliographystyle{elsart-num-sort}
\bibliography{dp}


\end{document}